\documentclass[fleqn,usenatbib]{mnras}

\usepackage[T1]{fontenc}
\usepackage{ae,aecompl}

\usepackage{graphicx}	
\usepackage{amsmath}	
\usepackage{amssymb}	
\usepackage{multirow}
\usepackage{pdflscape}
\usepackage{tabularx}
\newcolumntype{R}[1]{>{\raggedleft\arraybackslash}p{#1}}

\def\arcsec{\hbox{$^{\prime\prime}$}}

\newcommand{\wsc}{{\tt WSClean}}

\newcommand{\pybdsf}{{\tt pyBDSF}}
\newcommand{\trap}{{\tt TraP}}

\newcommand{\modp}{$V_{\nu}$}
\newcommand{\varp}{$\eta_{\nu}$}

\newcommand{\asec}{$^{\prime\prime}$}
\newcommand{\vardiff}{V$_{\mathrm{F}}$}

\newcommand{\pcc}{$r$}

\newcommand*\oline[1]{%
  \vbox{%
    \hrule height 0.5pt
    \kern0.25ex
    \hbox{%
      \kern-0.1em
      \ifmmode#1\else\ensuremath{#1}\fi
      \kern-0.1em
    }
  }
}

\usepackage{xcolor}

\newcommand{\anumba}{1}
\newcommand{\aname}{MKT\,J165945.1$-$484703}

\newcommand{\bnumba}{2}
\newcommand{\bname}{MKT\,J165955.1$-$491352}

\newcommand{\cnumba}{3}
\newcommand{\cname}{MKT\,J170028.1$-$482543}

\newcommand{\dnumba}{4}
\newcommand{\dname}{MKT\,J170057.2$-$484753}

\newcommand{\enumba}{5}
\newcommand{\ename}{MKT\,J170101.1$-$484953}

\newcommand{\fnumba}{6}
\newcommand{\fname}{MKT\,J170104.7$-$484842}

\newcommand{\gnumba}{7}
\newcommand{\gname}{MKT\,J170109.9$-$483550}

\newcommand{\hnumba}{8}
\newcommand{\hname}{MKT\,J170037.5$-$485646}

\newcommand{\inumba}{9}
\newcommand{\iname}{MKT\,J170145.8$-$484029}

\newcommand{\jnumba}{10}
\newcommand{\jname}{MKT\,J170154.7$-$485342}

\newcommand{\knumba}{11}
\newcommand{\kname}{MKT\,J170128.5$-$482955}

\newcommand{\lnumba}{12}
\newcommand{\lname}{MKT\,J170127.4$-$485810}

\newcommand{\mnumba}{13}
\newcommand{\mname}{MKT\,J170213.7$-$483337}

\newcommand{\gxnumba}{14}
\newcommand{\gx}{GX\,339$-$4}

\newcommand{\nnumba}{15}
\newcommand{\nname}{MKT\,J170225.5$-$485711}

\newcommand{\gxpsrnumba}{16}
\newcommand{\gxpsr}{PSR\,J1703$-$4851}

\newcommand{\onumba}{17}
\newcommand{\oname}{MKT\,J170355.9$-$485556}

\newcommand{\pnumba}{18}
\newcommand{\pname}{MKT\,J170340.2$-$484010}

\newcommand{\qnumba}{19}
\newcommand{\qname}{MKT\,J170404.0$-$485820}

\newcommand{\fbnumba}{20}
\newcommand{\fb}{MKT\,J170456.2$-$482100}

\newcommand{\rnumba}{21}
\newcommand{\rname}{MKT\,J170524.1$-$480842}

\newcommand{\snumba}{22}
\newcommand{\sname}{MKT\,J170546.3$-$484822}

\newcommand{\tnumba}{23}
\newcommand{\tname}{MKT\,J170721.9$-$490816}

\newcommand{\unumba}{24}
\newcommand{\uname}{MKT\,J170754.2$-$484252}

\emergencystretch=\maxdimen
\usepackage{newtxtext,newtxmath}

\title[21 new long-term variables in the \gx\,field]{21 new long-term variables in the \gx\,field: two years of MeerKAT monitoring}

\author[L. N. Driessen et al.]{L. N. Driessen,$^{1}$\thanks{E-mail: Laura@Driessen.net.au (LND)}
B. W. Stappers,$^{1}$
E. Tremou,$^{2}$
R. P. Fender,$^{3,4}$
P. A. Woudt,$^{3}$
\newauthor
R. Armstrong,$^{3,5}$
S. Bloemen,$^{6}$
P. Groot,$^{3,6,7}$
I. Heywood,$^{4,8}$
A. Horesh,$^{9}$
\newauthor
A. J. van der Horst,$^{10,11}$
E. Koerding,$^{6}$
V. A. McBride,$^{7,12,13}$
J. C. A. Miller-Jones,$^{14}$
\newauthor
K. P. Mooley$^{15,16}$
A. Rowlinson,$^{17,18}$
R. A. M. J. Wijers$^{17}$
\\ \\
$^{1}$Jodrell Bank Centre for Astrophysics, Department of Physics and Astronomy, The University of Manchester, Manchester, M13 9PL, UK\\
$^{2}$LESIA, Observatoire de Paris, CNRS, PSL Research University, Sarbonne Universite, Universite de Paris, Meudon, France\\
$^{3}$Inter-University Institute for Data Intensive Astronomy, Department of Astronomy, University of Cape Town, Private Bag X3,\\Rondebosch 7701, South Africa\\
$^{4}$Department of Physics, Astrophysics, University of Oxford, Denys Wilkinson Building, Keble Road, Oxford OX1 3RH, UK\\
$^{5}$South African Radio Astronomy Observatory, 2 Fir Street, Black River Park, Observatory, Cape Town 7925, South Africa\\
$^{6}$Department of Astrophysics/IMAPP, Radboud University, P.O. Box 9010, 6500 GL Nijmegen, The Netherlands\\
$^{7}$South African Astronomical Observatory, PO Box 9, Observatory 7935, South Africa\\
$^{8}$Department of Physics and Electronics, Rhodes University, PO Box 94, Makhanda 6140, South Africa\\
$^{9}$Racah Institute of Physics, The Hebrew University of Jerusalem, Jerusalem 91904, Israel\\
$^{10}$Department of Physics, The George Washington University, 725 21st Street NW, Washington, DC 20052, USA\\
$^{11}$Astronomy, Physics and Statistics Institute of Sciences (APSIS), 725 21st Street NW, Washington, DC 20052, USA\\
$^{12}$Department of Astronomy, University of Cape Town, Private Bag X3, Rondebosch 7701, South Africa\\
$^{13}$IAU Office of Astronomy for Development, Cape Town, 7935, South Africa\\
$^{14}$International Centre for Radio Astronomy Research -- Curtin University, GPO Box U1987, Perth, WA 6845, Australia\\
$^{15}$National Radio Astronomy Observatory, Socorro, NM 87801, USA\\
$^{16}$Caltech, 1200 E. California Blvd. MC 249-17, Pasadena, CA 91125, USA\\
$^{17}$Anton Pannekoek Institute, University of Amsterdam, Postbus 94249, 1090 GE, Amsterdam, The Netherlands\\
$^{18}$Netherlands Institute for Radio Astronomy (ASTRON), Oude Hoogeveensedijk 4, 7991 PD, Dwingeloo, The Netherlands\\
}
\date{Accepted 2022 March 14. Received 2022 February 17; in original form 2021 July 11}

\pubyear{2021}

\begin{document}
\label{firstpage}
\pagerange{\pageref{firstpage}--\pageref{lastpage}}
\maketitle

\begin{abstract}
We present 21 new long-term variable radio sources found commensally in two years of weekly MeerKAT monitoring of the low-mass X-ray binary \gx. The new sources vary on time scales of weeks to months {and have a variety of light curve shapes and spectral index properties.}
Three of the new variable sources are coincident with multi-wavelength counterparts; and one of these is coincident with an optical source in deep MeerLICHT images. For most sources, we cannot eliminate refractive scintillation of active galactic nuclei as the cause of the variability. {These new variable sources represent $2.2\pm0.5$\,per\,cent of the unresolved sources in the field, which is consistent with the 1-2\,per\,cent variability found in past radio variability surveys.} {However, we expect to find short-term variable sources in the field as well as these 21 new long-term variable sources.} We present the radio light curves and spectral index variability of the new variable sources, as well as the absolute astrometry and matches to coincident sources at other wavelengths. 
\end{abstract}



\section{Introduction}

We are entering a new era of radio  astronomy where we can execute {untargeted}, image-plane searches for variable and transient sources using sensitive instruments with wide-field capabilities. Instruments such as the Australian Square Kilometre Array Pathfinder
\citep[ASKAP\footnote{\href{https://www.atnf.csiro.au/projects/askap/index.html}{https://www.atnf.csiro.au/projects/askap/index.html}};][]{2021PASA...38....9H}, the Karl G. Jansky Very Large Array \citep[VLA;][]{perley2011}, the Low Frequency Array \citep[LOFAR;][]{2013A&A...556A...2V}, the Murchison Wide Field Array \citep[MWA;][]{2012rsri.confE..36T}, and the (more) Karoo Array Telescope \citep[MeerKAT;][]{2018ApJ...856..180C} are uncovering large samples of dynamic sources in the radio sky and facilitating their detailed light curve analyses without the need for targeting each source individually.

Previous surveys and investigations of the changing radio sky in the image-plane have revealed that $\sim$\,1--2\,per\,cent of radio point sources at L-band (1.4\,GHz) are variable \citep[see e.g.][for a review\footnote{See \href{http://www.tauceti.caltech.edu/kunal/radio-transient-surveys/index.html}{http://www.tauceti.caltech.edu/kunal/radio-transient-surveys/index.html} for an up-to-date list of {untargeted} radio surveys.}]{Ofek_2011}. Many of these past searches for variable sources used the VLA. For example,
\citet{VLA_Carilli} searched the Lockman Hole region with the VLA on time scales of 19\,days to 17\,months and found that less than 2\,per\,cent of sources varied;
\citet{VLA_de_Vries} used Faint Images of the Radio Sky at Twenty Centimetres \citep[FIRST;][]{FIRST1995} observations of the zero-Declination strip and found that $\sim$2\,per\,cent of sources varied; \citet{VLA_Levinson} searched VLA FIRST and National Radio Astronomy Observatory (NRAO) VLA Sky Survey \citep[NVSS;][]{NVSS1998} observations and found one transient candidate; and \citet{2013ApJ...768..165M} observed the Extended Chandra Deep Field South with the VLA at 1.4\,GHz and found that 1\,per\,cent of unresolved sources were variable.
The Nasu Sky Survey detected one confirmed transient source, supernova WJN\,J1443$+$3439 \citep{NASU_Transients,2007PASP..119..122K,2007ApJ...657L..37N,2007AJ....133.1441M,2008NewA...13..519K,2009AJ....138..787M,2009ApJ...704..652N}.
Recently, the VLA COSMOS HI Large Extragalactic Survey \citep[CHILES;][]{2013ApJ...770L..29F,2016ApJ...824L...1F} field was observed 172 times over 5.5\,years by the CHILES Variable and Explosive Radio Dynamic Evolution Survey \citep[CHILES VERDES;][]{2020arXiv200905056S} team. This was a very sensitive search for variable and transient sources, with each epoch reaching a root-mean-square (RMS) noise of $\sim10\,\mathrm{\mu Jy}$. 
However, they only used the VLA while it was in B-configuration, which occurs for approximately four months per sixteen month cycle. During those four months, CHILES VERDES observed the field for 1-8\,hours per epoch every one to two days. They then had a twelve month break where they did not monitor their sources.
The CHILES VERDES survey has unprecedented depth and cadence, revealing that 58 of their 2713 ($\sim$2\,per\,cent) sources are variable. 
Other telescopes have also been used to search for variables, such as the Molonglo Observatory Synthesis Telescope \citep[MOST;][]{1981PASAu...4..156M} and ASKAP.
\citet{2011MNRAS.412..634B} used MOST to observe a large part of the sky over 22\,years. They found variability in less than 0.3\,per\,cent of the radio sources they observed.
The ASKAP Boolardy Engineering Test Array \citep[BETA;][]{2014PASA...31...41H} was used to searched for variable sources in two regions. BETA consisted of 6 ASKAP antennas, and
only one variable source was found with this instrument \citep{2016MNRAS.456.3948H,2016MNRAS.457.4160H}. \citet{2021arXiv210106048W} used the full ASKAP array to search for intra-hour variable sources and they found 6 variable sources.

These surveys and searches for variable sources tell us that only a small percentage of the radio sky is variable. However, we know that those radio sources that are variable reveal important information about some of the most extreme and explosive astrophysical sources. This includes black-hole accretion and jets in X-ray binaries \citep[e.g.][]{2020MNRAS.493L.132T}, gamma-ray burst afterglows \citep[e.g.][]{2014PASA...31....8G,2016AdAst2016E..16C}, and jet-shocks from Active Galactic Nuclei \citep[AGN; e.g.][]{2008A&A...485...51H}. Radio waves are not obstructed by dust and gas, meaning that radio observations can be used to obtain accurate rates of events such as core-collapse supernovae, tidal disruption flares, and possibly type Ia supernovae. As well as intrinsic effects there are extrinsic effects, for example five of the variable sources found by \citet{2021arXiv210106048W} line up on the sky, suggesting scintillation by an interstellar medium (ISM) filament. Expanding {untargeted} searches for radio variables and transients could also reveal new classes of transients.

It is thought that AGN variability is the dominant source of radio point source variability \citep[e.g.][]{VLA_Thyagarajan,Mooley_sourcematching}. AGNs are observed to vary on short timescales of hours to days, thought to be due to refractive interstellar scintillation \citep[e.g.][]{1990ARA&A..28..561R}. Refractive interstellar scintillation is caused by electrons along the line of sight, and the time scale and variability amplitude caused by scintillation depends on the Galactic latitude of the AGN \citep[][]{2019arXiv190708395H}. AGN are observed to flare on longer timescales, from days to months, due to shocks in the jets formed by material accreting onto the black holes at their centres \citep{2008A&A...485...51H}. In the CHILES VERDES survey, \citet{2020arXiv200905056S} determined that most of their variable sources were AGN using their multi-wavelength counterparts and radio spectral indices. Compact AGN are expected to have flat spectra, $\alpha\gtrsim-0.5$ \citep[e.g.][and references therein]{2017A&ARv..25....2P}, where the spectral index, $\alpha$, is given by $S_{\nu}\propto\nu^{\alpha}$ where $S_{\nu}$ is the flux density of the source at frequency $\nu$.

MeerKAT \citep{2018ApJ...856..180C} is a 64-dish interferometer in the Karoo region of South Africa. Each dish has an effective collecting area  with a diameter of 13.5\,m and the longest baseline is 8\,km, giving a resolution of $\sim5$\arcsec and a field of view (FoV) of $\sim$1\,square\,degree at 1400\,MHz. ThunderKAT\footnote{The HUNt for Dynamic and Explosive Radio transients with MeerKAT} is a MeerKAT large survey project (LSP) investigating variable and transient radio sources in the image plane, including commensal searches \citep{2017arXiv171104132F}. ThunderKAT has committed to observing the low-mass X-ray binary \gx\,on a weekly cadence for five years, beginning in September 2018 \citep{2020MNRAS.493L.132T}. This makes the \gx\,field ideal for commensal searches for variable and transient sources, with the first MeerKAT transient, \fb, discovered in this field \citep[][]{FlareyBoi}.

We present the results of searching for {long-term (sources that show variability on scales of weeks to months) variable sources} in the \gx\,field over the first two years of ThunderKAT monitoring of the source. In Section\,\ref{sec: VAR MKT obs} we present our MeerKAT observations, and in Section\,\ref{sec: VAR source matching} we present our method for matching sources to their multi-wavelength counterparts. In Section\,\ref{sec: VAR results} we present our results and in Sections\,\ref{sec: VAR discussion} and \ref{sec: VAR conclusions} we discuss and conclude.

\section{MeerKAT radio observations}
\label{sec: VAR MKT obs}

We present 102 epochs of weekly MeerKAT observations of the field surrounding \gx. ThunderKAT started weekly monitoring of the field in September 2018 \citep{2020MNRAS.493L.132T} using the L-band (856-1712\,MHz) receiver in full polarisation mode. The MeerKAT L-band receiver has a bandwidth of 856\,MHz, a central frequency of 1284\,MHz, and 4096 frequency channels. The field is observed for $\sim$10\,minutes each week with a minimum integration time of 8\,seconds. The phase calibrator (1722$-$554) is observed for 2\,minutes before and after observing the target field, and the band-pass and flux calibrator (1934$-$638) is observed for 5\,minutes at the start of the observing block.

Full details on the processing of the weekly \gx\,observations can be found in \citet{FlareyBoi} and \citet{2020MNRAS.493L.132T}. The data are flagged using 
{\tt AOFlagger}\footnote{\href{https://aoflagger.readthedocs.io/en/latest/index.html}{https://aoflagger.readthedocs.io/en/latest/index.html}}
\citep{2010MNRAS.405..155O,2012A&A...539A..95O} and are calibrated using the Common Astronomy Software Application\footnote{\href{https://casa.nrao.edu/}{https://casa.nrao.edu/}}  \citep[CASA;][]{CASA}. Calibration includes phase correction, antenna delays and band-pass corrections. 
The data are imaged using \wsc\,\citep{offringa-wsclean-2014}, including $w$-projection planes, a Briggs robust weighting of $-$0.7 \citep[][]{1995AAS...18711202b}, and multi-scale clean.
The multi-frequency synthesis (MFS) images were produced using 8 frequency channels and a 4$^{\mathrm{th}}$ order spectral polynomial fit. {The weekly, ten-minute MFS images have a typical RMS noise of $\sim30\,\mathrm{\mu Jy\,beam^{-1}}$.} We produce 8 subband images per epoch by excluding the {\tt -join-channels} parameter and spectral fit. The 8 subbands have central frequencies: 909, 1016, 1123, 1230, 1337, 1444, 1551, 1658\,MHz and each subband image is primary beam corrected. The primary beam correction is performed by multiplying the final fits image by the primary beam model for each subband. The 1230\,MHz subband is strongly affected by radio frequency interference (RFI), and as such we exclude it from our analysis. As we are focusing on long-term variability in this investigation, we only produce full time-integration images, both subband and MFS, for each epoch. We will present an investigation of the short-term variability of sources in the \gx\,field in a future publication.
As well as the single-epoch images, we also utilise a deeper, combined image of the field produced using {\tt DDFacet} \citep{2018tasse} to determine the positions of the sources. This image was produced by jointly imaging the visibilities from 8 epochs (September and October 2018, and a commissioning image from 2018 April) with a total integration time of 3.63\,hours.

\subsection{The LOFAR Transients Pipeline}
\label{sec: VAR trap}

The {\tt LOFAR Transient Pipeline} \citep[\trap, Release 4.0;][]{Swinbank2015} is a software package for extracting light curves from a time series of fits images. It has been designed with radio images in mind, specifically to find sources in LOFAR images for the Transients Key Project\footnote{\href{https://transientskp.org/}{https://transientskp.org/}}.

We used the default \trap\,parameters with some minor adjustments to search the \gx\, field for variable and transient sources. The default settings for the pipeline configuration and job configuration files can be found in the \trap\,documentation\footnote{\href{https://tkp.readthedocs.io/en/latest/userref/config/}{https://tkp.readthedocs.io/en/latest/userref/config/}}.
We used the default signal-to-noise (S/N) of $8$ for a new source to be detected in an image, and we set the {\tt force\_beam} parameter to True (default is False) to search for sources with a Gaussian shape consistent with the shape of the synthesised beam in each image.
{The {\tt beamwidths\_limit} parameter was set to $3.0$. This means that there must be at least three synthesised beamwidths between two sources for those two sources to be considered unique.}
Both {\tt force\_beam} and {\tt beamwidths\_limit} were set this way to reduce the number of extended sources, particularly double-lobed galaxies, detected as point sources.
For our statistics and information we require flux density measurements in every epoch, even if those measurements are upper limits. As such, we force \trap\,to continue measuring flux densities for all sources in all epochs by setting the {\tt expiration} parameter to $150$. \trap\,searches through the inputted images in chronological order. So, as mentioned above, only a source detected in the first epoch will be tracked in all epochs. To track as many sources as possible for as many epochs as possible, we insert the deep MeerKAT image of the \gx\,field as the ``first epoch''. Once \trap\,has run, we remove this epoch from our analysis. {Using the deep image as our ``first epoch'' and a \trap\,detection threshold of 8 means that the tracked sources have a minimum S/N of 8 in this deep image, but may have a much lower S/N in the weekly ten-minute epochs. As such, future mentions of S/N for sources in this work are calculated using the simple method of dividing the measured flux density by the uncertainty on the flux density.}
We extract the source information and light curves from \trap\, using {\tt Python}\footnote{The code for this can be found on GitHub: \href{https://doi.org/10.5281/zenodo.4456303}{https://doi.org/10.5281/zenodo.4456303}}.

\subsection{Variability parameters}
\label{sec: VAR var params}

The \trap\, {software calculates} two parameters, \varp\,and \modp, to determine which sources are variable.
The \varp\,parameter is based on the reduced $\chi^{2}$ statistic:
\begin{eqnarray}
    \eta_{\nu} & = & \frac{1}{N-1}\sum^{N}_{i=1}\frac{\left(I_{\nu,i}-\oline{I_{\nu}}\right)}{\sigma^{2}_{\nu,i}} \nonumber \\
     & = & \frac{N}{N-1}\left(\oline{wI^{2}} - \frac{\oline{wI}^{2}}{\oline{w}} \right)
    \label{eq: VAR chi2 param}
\end{eqnarray}
where {$N$ is the number of measurements, $I_{\nu,i}$ is the flux density at frequency $\nu$ and epoch $i$, $\sigma_{\nu,i}$ is the uncertainty on $I_{\nu,i}$, and $w$ is the weight ($w_{i}=1/\sigma_{\nu,i}^{2}$)}. A source with a low \varp\,value is consistent with a constant source, and a high \varp\,means that the source deviates from a constant source.
The \modp\,parameter is defined by:
\begin{eqnarray}
    V_{\nu} & = & \frac{s}{\oline{I_{\nu}}} \nonumber \\
     & = & \frac{1}{\oline{I_{\nu}}}\sqrt{\frac{N}{N-1}\left(\oline{I_{\nu}^2}-\oline{I_{\nu}}^2 \right)} \label{eq: VAR Mod Param}
\end{eqnarray}
{where $\oline{I_{\nu}}$ and $s$ are the light curve mean and standard deviation respectively.}
Sources with a low \modp\,have a smaller spread of flux densities while a high \modp\,indicates a larger spread of flux densities and hence variability. It is important to note that the \modp\,parameter does not include uncertainties, which can lead to low S/N sources appearing variable.
We use both \varp\,and \modp\,to investigate sources in the \gx\,field.

\subsection{Light curve binning}
\label{sec: binning method}

As we are interested in the long-term variability of the sources in the field and we wanted to confirm longer term low amplitude variability, we binned the light curves for all sources into ten-epoch bins {using the weighted mean}.
We then calculated the variability parameters for each source using the binned light curves. 
To ensure that any variability was not 
{dependent on the starting epoch, we performed the same analysis after removing epoch one, epochs one and two, and epochs one, two and three.}
This did not impact the light curves or variability parameters.
{We found that binning enhances the \varp\,parameter for light curves where the binning is on comparable time scales to the trend in the curve. This is because a trend across multiple bins enhances the trend, as opposed to binning randomly scattered data.}

{We used the binned flux density values to calculate the variability indicator\,\citep[\vardiff, similar to \vardiff\,as defined by;][]{Ofek_2011} for each source, using the weighted mean ($\oline{I_{\nu}}$), minimum ($I_{\nu,\mathrm{min}}$), and maximum ($I_{\nu,\mathrm{max}}$) flux density values:
\begin{eqnarray}
    D = \frac{I_{\nu,\mathrm{max}} - I_{\nu,\mathrm{min}}}{\oline{I_{\nu}}}\,.
\end{eqnarray}
The weighted mean is given by:
\begin{eqnarray}
    \oline{I_{\nu}}=\frac{\sum^N_{i=1}\left(I_{\nu,i}\times w_{\nu,i}\right)}{\sum^N_{i=1}w_{\nu,i}}
\end{eqnarray}
with uncertainty:
\begin{eqnarray}
    \sigma_{\mathrm{mean}}=\frac{1}{\sqrt{\sum^N_{i=1}w_{\nu,i}}}
\end{eqnarray}
where the weights are given by $w=\sigma_{\nu,i}^{-2}$ where $\sigma_{\nu,i}$ is the uncertainty on the $i^{\mathrm{th}}$ flux density $I_{\nu,i}$. The uncertainty on $D$ is then given by:
\begin{eqnarray}
    \sigma_D = \oline{I_{\nu}} \times \sqrt{\left( \frac{\sigma_{\mathrm{mean}}}{\oline{I_{\nu}}} \right)^2 + \left( \frac{\sqrt{\sigma_{\nu,\mathrm{max}}^2 + \sigma_{\nu,\mathrm{min}}^2}}{I_{\nu,\mathrm{max}} - I_{\nu,\mathrm{min}}} \right)^2} \,.
\end{eqnarray}
We then multiply $D\pm\sigma_D$ by 100 to obtain the percentage \vardiff.}

\subsection{Systematic effect corrections}
\label{sec: VAR sys effects}

In the initial stages of investigating the sources in the \gx\,field, we manually examined the light curves of all of the sources. We noticed that many 
had similar underlying light curve shapes, and that many more sources than expected appeared to be variable. We determined that some of the light curve correlations between unrelated sources are due to those sources being small (only slightly larger than the synthesised beam), resolved sources. We removed the resolved sources and spatially close sources from the data set reducing the number and strength of the correlations between sources {(see Appendix\,\ref{app: systematics})}; however, some correlation between light curves remained. The remaining underlying systematic effects are multiplicative and cause variations of $\sim$10\,per\,cent.

We determined the shape of the systematics by taking all of the sources with a S/N$\geq$3 and dividing their flux densities by the flux density in a reference epoch. {We used this S/N threshold as we needed a sufficient number of sources to identify the systematics and to test any possible flux density and position dependence. We chose the last epoch as the reference epoch} as all sources have a measurement in the last epoch due to the forced measurements (see Section\,\ref{sec: VAR trap}). We then take the distribution of the scaled flux density of every source from each epoch and find the median and the median absolute deviation (MAD). 
These median values (with the MAD as the uncertainty) are the model of the systematics for each epoch.

To correct the light curve of each source for the systematics, we divide each light curve by the model and propagate the uncertainties. We then recalculate \varp\,and \modp\,for each source using the corrected flux density and Equations\,\ref{eq: VAR chi2 param} and \ref{eq: VAR Mod Param}. The light curves and variability parameters discussed in this paper have all been corrected for the systematic effects.
{These light curve systematics are the reason that we have chosen to focus on long-term variable sources instead of short-term variable sources. The systematics particularly induce week-to-week variability, as opposed to longer trends.}
For more information regarding the systematics and corrections, see Appendix\,\ref{app: systematics}. 

\subsection{Absolute astrometry}
\label{sec: VAR absolute astrometry}

We performed a Python Blob Detector and Source Finder\footnote{\href{https://www.astron.nl/citt/pybdsf/}{https://www.astron.nl/citt/pybdsf/}} (\pybdsf) search on the deep MeerKAT image to determine the source positions and found 17130 sources. The \pybdsf\,software is a source extractor designed for LOFAR and is focused on cleanly extracting all flux from an image.
The maximum uncertainties for the Right Ascension and Declination positions of our sources of interest are 0\farcs08 and 0\farcs09 respectively. To accurately match the sources to sources from other catalogues, we need to understand the accuracy of our absolute astrometry.

There are no matches between MeerKAT sources in our FoV and the third International Celestial Reference Frame \citep[3ICRS;][]{2020A&A...644A.159C}. However, there are 11 Australian Telescope Compact Array (ATCA) Parkes-MIT-NRAO (PMN) source matches \citep[ATPMN;][]{2012MNRAS.422.1527M}. ATPMN is a source catalogue where 8385 PMN sources were observed with ATCA at 8.6\,GHz, resulting in a catalogue of 9040 radio sources. The number of sources from ATPMN is greater than the number of PMN sources as the higher resolution of ATCA resolves some PMN sources into multiple sources. \citet{2012MNRAS.422.1527M} compared the ATPMN sources to the positions of the Long-Baseline Array (LBA) Calibrator Survey 1 catalogue of Southern Sources \citep[LCS1;][]{2011MNRAS.414.2528P} and to the International Celestial Reference Frame \citep[ICRF;][]{1998AJ....116..516M}. They matched ATPMN sources to 309 LCS1 and 26 ICRF sources, and found that the median astrometric uncertainty in the ATPMN positions is 0\farcs4 in both Right Ascension and Declination.

Of the 11 ATPMN sources within the MeerKAT \gx\,FoV, 5 are resolved in the MeerKAT observations. As resolved sources have poorer localisation precision, we discard these 5 sources.
We use the 6 remaining ATPMN sources, shown in Table\,\ref{tab: ATPMN source info}, and the corresponding 6 MeerKAT sources to test and correct the astrometry.

\begin{table*}
    \centering
    \begin{tabular}{lrrrrr}
    ATPMN name & RA & DEC & $\mathrm{S_{5\,GHz}}$ (mJy) & $\mathrm{S_{8\,GHz}}$ (mJy) & $\mathrm{\alpha}$ \\
\hline
\hline
J165418.2$-$481303 & 253.5759 & $-$48.2176 & $82\pm7$  & $61\pm10$ & $-0.5\pm0.3$ \\
J165613.1$-$492318 & 254.0549 & $-$49.3883 & $110\pm7$ & $46\pm10$ & $-1.5\pm0.4$ \\
J165614.9$-$472915 & 254.0623 & $-$47.4876 & $77\pm7$  & $37\pm10$ & $-1.3\pm0.5$ \\
J165902.0$-$474618 & 254.7585 & $-$47.7719 & $61\pm7$  & $71\pm10$ & $0.3\pm0.3$ \\
J165908.3$-$481548 & 254.7848 & $-$48.2635 & $99\pm7$  & $43\pm10$ & $-1.4\pm0.4$ \\
J171154.9$-$491250 & 257.9791 & $-$49.2141 & $73\pm7$  & $43\pm10$ & $-0.9\pm0.4 $\\
\hline
    \end{tabular}
    \caption[Summary of ATPMN sources.]{Summary of the ATPMN sources used to determine the absolute astrometry of the MeerKAT observations of the \gx\,field. The Right Ascension (RA) and Declination (DEC) are given in degrees.  The $\mathrm{\alpha}$ value is the ATPMN spectral index.}
    \label{tab: ATPMN source info}
\end{table*}

We fit for a transformation matrix to shift and rotate the MeerKAT deep image source positions to match the 6 ATPMN source positions. We then apply the transformation to all MeerKAT sources in the field\footnote{The code for performing the astrometric corrections can be found on GitHub: \href{https://doi.org/10.5281/zenodo.4921715}{https://doi.org/10.5281/zenodo.4921715}}. 
The separation between the MeerKAT and ATPMN reference sources before and after transformation are shown in Table\,\ref{tab: astrometry sep}. To determine the uncertainties on the transformed MeerKAT positions we performed a Monte Carlo simulation. We selected a position for each ATPMN reference source from a Gaussian with the mean of the ATPMN position and a standard deviation of 0\farcs4. We then calculated and applied the transformation matrix to the MeerKAT reference sources. We repeated these steps 5000 times, and found the standard deviation on the transformed positions of the MeerKAT reference sources. The minimum and maximum standard deviations were 0\farcs2 and 0\farcs4 for both the Right Ascension and Declination. This uncertainty is much larger than the uncertainty derived by \pybdsf\,for our sources of interest, and as such we will use an uncertainty of 0\farcs4 in Right Ascension and Declination for all of our MeerKAT sources.
The final positions we use for our MeerKAT sources are therefore the transformed position from the deep MeerKAT stack, with an uncertainty of 0\farcs4 and this means that the positions for the sources quoted in this paper have been corrected for absolute astrometry.

\begin{table}
    \centering
    \begin{tabular}{lrr}
    ATPMN source name & separation before (\asec) & separation after (\asec) \\
    \hline
    \hline
    J165418.2$-$481303 & 1.22 & 0.11 \\
    J165613.1$-$492318 & 0.86 & 0.05 \\
    J165614.9$-$472915 & 1.34 & 0.18 \\
    J165902.0$-$474618 & 0.85 & 0.20 \\
    J165908.3$-$481548 & 0.88 & 0.09 \\
    J171154.9$-$491250 & 0.46 & 0.06 \\
    \hline
    \end{tabular}
    \caption[Separation between ATPMN and MeerKAT reference sources before and after transformation.]{Separation between ATPMN and MeerKAT reference sources before and after applying the transformation. The separation is given in arcseconds.}
    \label{tab: astrometry sep}
\end{table}

\section{Source matching}
\label{sec: VAR source matching}

The corrections to the absolute astrometry in the common reference frame allow us to match sources to objects in other catalogues at different frequencies. We used {\tt Astropy Astroquery}\footnote{\href{https://astroquery.readthedocs.io/en/latest/}{https://astroquery.readthedocs.io/en/latest/}} to search for sources in Vizier \citep{vizierpaper} that are co-located with our sources of interest. We searched the Vizier catalogues shown in Table\,\ref{tab: vizier cats} using the astrometric precision for each catalogue. If a catalogue has a smaller astrometric uncertainty than our positions we searched within a radius of 0\farcs4, otherwise we used the radius defined by the catalogue's uncertainty. Known variable sources \gx\,and \fb\,match with multiple catalogues, as expected.

We use this same method to match source positions to MeerLICHT \citep{10.1117/12.2232522} and the Rapid ASKAP Continuum Survey \citep[RACS;][]{2020PASA...37...48M}. MeerLICHT (more light) is a fully robotic, 0.65\,metre optical telescope at the Sutherland station of the South African Astronomical Observatory (SAAO). MeerLICHT has a 2.7\,square\,degree FoV. 
A deep image using 15 1-minute epochs of the \gx\,field has been made in 5 SDSS filters ($u$, $g$, $r$, $i$, $z$) and the wider $q$ filter (440-720\,nm).
The individual epochs are being reprocessed as part of a pipeline update, but we can match our MeerKAT sources to MeerLICHT detections in the deep images. 
{The MeerLICHT coordinates are in the International Celestial Reference System (ICRS) using the Gaia Data Release 2 \citep[DR2, J2015.5;][]{2016A&A...595A...1G,2018A&A...616A...1G} frame. The coordinates are consistent with FK5 J2000 within 0\farcs1.}
RACS is a radio survey covering the whole southern sky (Declinations below +41\,degrees) with ASKAP. The central frequency is 887.5\,MHz with a bandwidth of 288\,MHz, and the median RMS in each image is 0.25\,mJy\,per\,beam. The resolution is approximately 15\asec. The bottom of the MeerKAT band is 856\,MHz, so there is an overlap between RACS and MeerKAT. The RACS data were released in early 2021, and can be found online\footnote{\href{https://research.csiro.au/racs/home/survey/}{https://research.csiro.au/racs/home/survey/}}. The catalogues of sources from RACS provide the position, flux density and spectral index of the sources. The RACS positions have not been corrected for absolute astrometry and the spectral indices do not include uncertainties.
The RACS synthesised beam is more than 3 times larger than the MeerKAT synthesised beam, which means that some resolved MeerKAT sources or sources that are close to each other will appear as one source in the RACS images. We therefore match the MeerKAT sources to the RACS sources by finding the minimum separation between sources, and then confirm the matches by visual inspection, checking that each RACS source is only matched to a single MeerKAT source.

\section{Results}
\label{sec: VAR results}

\begin{figure}
\includegraphics[width=\columnwidth]{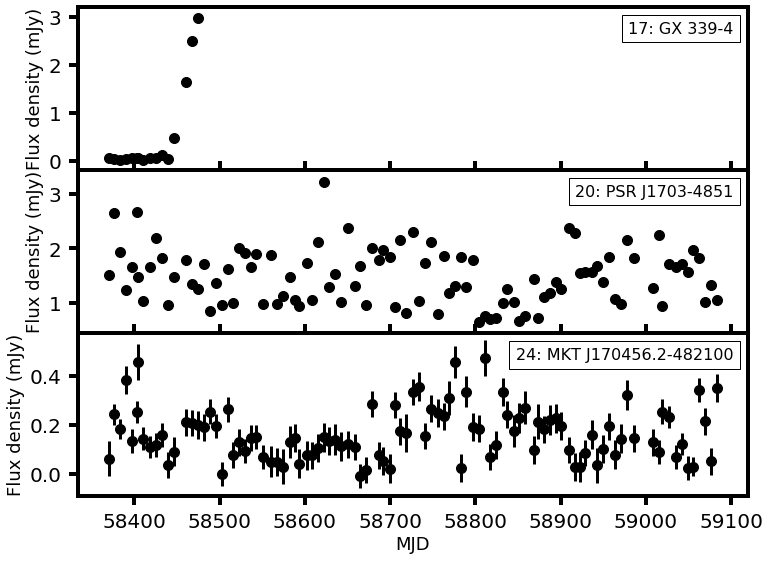}
\caption[MFS light curves of known variable sources \gx, \fb, and \gxpsr.]{MFS light curves of known variable sources \gx, \gxpsr, and \fb. We have included the light curve of \gx\,as shown in \citet{2020MNRAS.493L.132T}, see Tremou et al. (2021, in prep.) for an up-to-date light curve.}
\label{fig: VAR MFS GX FB PSR}
\end{figure}

\begin{figure*}
\includegraphics[width=0.9\textwidth]{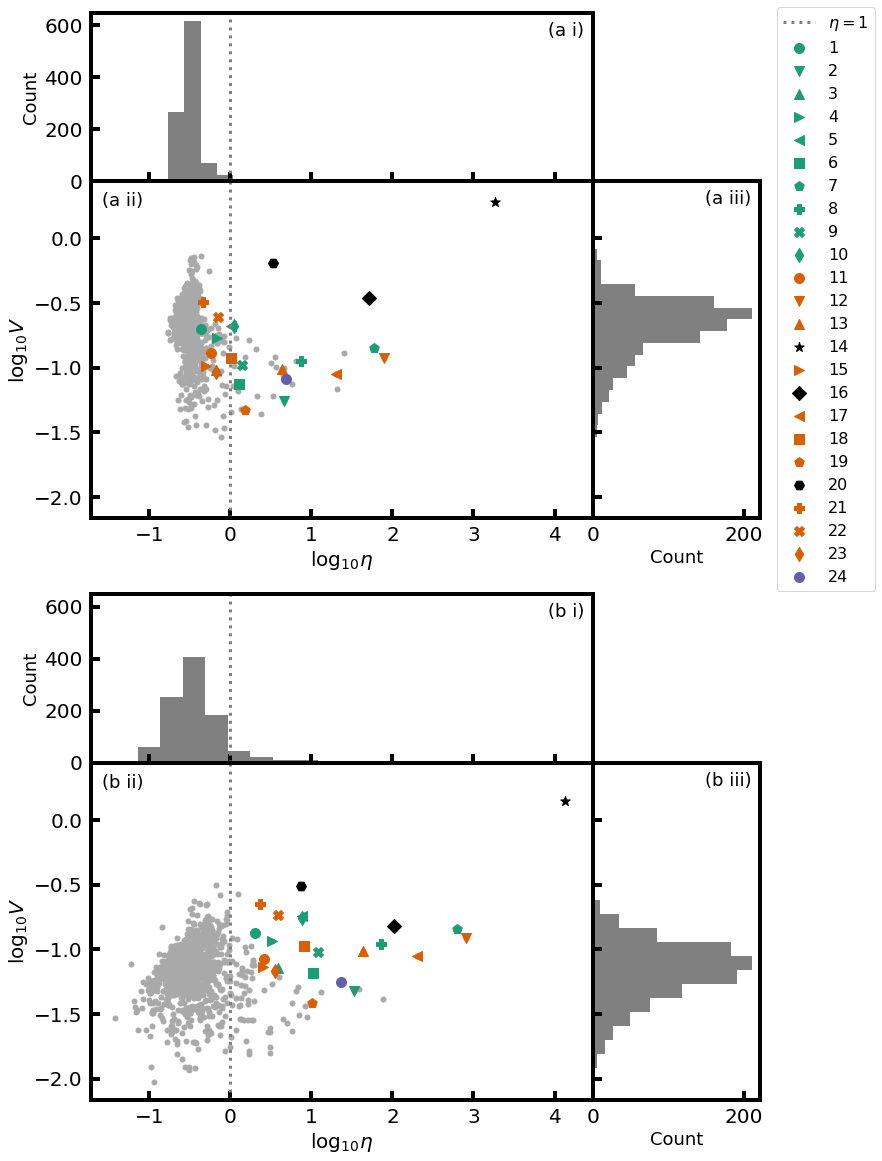}
\caption{{Variability parameters for sources in the \gx\,field calculated using the MFS light curves.
Panel (a ii) shows the variability parameters for all sources in the field (grey markers) and the variable sources using the original, unbinned light curves. 
Panels (a i) and (a iii) show the distributions of the unbinned $\eta_{\mathrm{MFS}}$ and $V_{\mathrm{MFS}}$ parameters respectively.
Panel (b ii) shows the variability parameters for all sources in the field (grey markers) and the variable sources using the ten-epoch binned light curves. 
Panels (b i) and (b iii) show the distributions of the binned $\eta_{\mathrm{MFS}}$ and $V_{\mathrm{MFS}}$ parameters respectively.
The grey-dashed line in panels (a i), (a ii), (b i) and (b ii) indicates where $\eta_{\mathrm{MFS}}=1$ (or $\log_{10}\eta_{\mathrm{MFS}}=0$).
All of the sources in both panels have been detected with a S/N of three in at least one epoch.
The numbers identifying the long-term variable sources in the legend are in Table\,\ref{tab: VAR supp source spec summary}.
We note that there are more outliers in the variability parameters than the long-term variable sources that we discuss in this paper.
These sources are short-term (week-to-week variability) variable source candidates and will be discussed in future work.}
}
\label{fig: VAR var params}
\end{figure*}

We detect 1080 unique point sources at least once with S/N$>3$ in the weekly, ten-minute images of the \gx\,field. The light curves of known variable sources \gx\,\citep{2020MNRAS.493L.132T}, \fb\,\citep[][]{FlareyBoi}, and 
{mode-changing pulsar  \gxpsr\,\citep[][]{2007MNRAS.377.1383W,2019MNRAS.484.3691J}} are shown in Figure\,\ref{fig: VAR MFS GX FB PSR}. While we will not be investigating these sources in depth in this paper, we include them to demonstrate the light curves of variable sources and to demonstrate the variability of outlier sources in the variability parameter plots, Figure\,\ref{fig: VAR var params}. {In this work we investigate only long-term variable sources, sources that vary on time scales of months or more. There are other outlier sources in Figure\,\ref{fig: VAR var params}, these sources are short-term variable candidates and will be further investigated in future work.} The light curve of \gx\,up to early 2021 will be discussed in Tremou et al. (in prep).

\subsection{New long-term variable radio sources}
\label{sec: VAR results new sources}

We found 21 new long-term variable sources\footnote{The light curve data for these 21 new variable sources can be found here: \href{http://doi.org/10.5281/zenodo.5069119}{http://doi.org/10.5281/zenodo.5069119}} within one degree of \gx\,using a combination of manual vetting and the \varp\,and \modp\,variability parameters. Manual vetting was performed by looking at the light curves for every source in all seven subbands and MFS for every source with a S/N\,$>3$ in at least one epoch (1080 unique sources). The median flux density of each source was plotted together with the light curve, and deviation from the median as well as peak-to-peak variation was used to identify variable sources.
{We initially plotted and manually looked through all unbinned light curves to investigate the systematic variability (see Section\,\ref{sec: VAR sys effects}), and while performing this investigation we noted the sources that appeared to vary over the long-term.
This is why many of the new long-term variable sources were found via manual vetting, while a few were found as they are extreme outliers in the unbinned \varp\,parameter.
After binning (see Section\,\ref{sec: binning method}) we found that all of the sources that were identified manually have a binned \varp\,value $>1$, shown in Figure\,\ref{fig: VAR var params} and Table\,\ref{tab: VAR supp source spec summary}, and would have been identified as candidates using this method.}
A summary table of the positions of the variable sources in the \gx\,field is also shown in Table\,\ref{tab: VAR supp source spec summary}. We will refer to these sources by their MeerKAT names.  Upon visual inspection of the in-band spectra, we assume that the spectral index can be modelled by a power-law for all of the long-term variable sources. We use non-linear least squares fitting to fit a power-law to the 7 subband flux densities for each source per epoch. The weighted mean MeerKAT spectral indices for each source are shown in Table\,\ref{tab: VAR supp source spec summary}.

\begin{table*}
    \centering
\begin{tabular}{llrrrrR{1.8cm}R{1.6cm}}
 & Name & $\oline{\alpha}$ & \vardiff\,(\%) & S$_{909\,\mathrm{MHz}}$ (mJy) & S$_{\mathrm{RACS}}$ (mJy) & $\eta_{\mathrm{MFS}}$/$V_{\mathrm{MFS}}$ (unbinned) & $\eta_{\mathrm{MFS}}$/$V_{\mathrm{MFS}}$ (binned) \\
\hline
\hline
 \anumba & \aname & $0.19\pm0.05$ & $45\pm13$ & $0.29\pm0.03$ &  & 0.43/0.2 & 2.0/0.13 \\
 \bnumba & \bname & $-0.44\pm0.009^{*}$ & $12\pm1$ & $9.13\pm0.05$ & $9.81\pm0.06$ & 4.5/0.055 & 34/0.048 \\
\cnumba & \cname & $-0.24\pm0.02$ & $21\pm5$ & $1.10\pm0.03$ &  & 0.67/0.096 & 3.9/0.072 \\
\dnumba & \dname & $0.5\pm0.05$ & $32\pm9$ & $0.26\pm0.02$ &  & 0.69/0.17 & 3.3/0.11 \\
\enumba & \ename & $0.54\pm0.05$ & $52\pm9$ & $0.25\pm0.02$ &  & 1.0/0.21 & 7.7/0.18 \\
\fnumba & \fname & $-0.11\pm0.02$ & $20\pm3$ & $1.13\pm0.03$ & $1.96\pm0.06$ & 1.3/0.074 & 11/0.065 \\
\gnumba & \gname & $1.3\pm0.01^{*}$ & $47\pm1$ & $3.19\pm0.03$ & $3.32\pm0.05$ & 59/0.14 & 620/0.14 \\
\hnumba & \hname & $0.95\pm0.01^{*}$ & $31\pm2$ & $1.81\pm0.03$ & $1.51\pm0.03$ & 7.5/0.11 & 73/0.11 \\
\inumba & \iname & $-0.58\pm0.02^{*}$ & $26\pm4$ & $0.90\pm0.02$ &  & 1.4/0.1 & 12/0.095 \\
\jnumba & \jname & $0.43\pm0.05$ & $46\pm9$ & $0.23\pm0.02$ &  & 1.1/0.21 & 7.7/0.17 \\
\knumba & \kname & $0.38\pm0.04$ & $26\pm8$ & $0.38\pm0.03$ &  & 0.57/0.13 & 2.6/0.084 \\
\lnumba & \lname & $-0.023\pm0.01^{*}$ & $35\pm1$ & $13.62\pm0.05$ & $14.3\pm0.2$ & 79/0.12 & 805/0.12 \\
\mnumba & \mname & $0.015\pm0.01^{*}$ & $26\pm2$ & $1.61\pm0.03$ &  & 4.3/0.097 & 43/0.098 \\
\gxnumba & \gx & $0.079\pm0.01^{*}$ & $1268\pm7$ & $2.28\pm0.03$ & $1.49\pm0.04$ & 1845/1.9 & 13655/1.4 \\
\nnumba & \nname & $-0.19\pm0.03$ & $21\pm7$ & $0.48\pm0.02$ &  & 0.5/0.1 & 2.5/0.073 \\
\gxpsrnumba & \gxpsr & $-2.0\pm0.01$ & $57\pm2$ & $3.09\pm0.03$ & $4.11\pm0.1$ & 52/0.34 & 106/0.15 \\
\onumba & \oname & $0.31\pm0.01^{*}$ & $31\pm1$ & $3.98\pm0.03$ & $4.73\pm0.06$ & 20/0.09 & 200/0.089 \\
\pnumba & \pname & $-0.14\pm0.03$ & $30\pm5$ & $0.54\pm0.02$ &  & 1.0/0.12 & 8.1/0.11 \\
\qnumba & \qname & $-0.49\pm0.01^{*}$ & $13\pm2$ & $2.80\pm0.03$ & $2.96\pm0.03$ & 1.5/0.047 & 10/0.038 \\
\fbnumba & \fb & $1.0\pm0.06$ & $90\pm17$ & $0.19\pm0.03$ &  & 3.3/0.65 & 7.4/0.31 \\
\rnumba & \rname & $1.1\pm0.06$ & $70\pm20$ & $0.45\pm0.04$ &  & 0.46/0.32 & 2.3/0.22 \\
\snumba & \sname & $-1.5\pm0.05$ & $53\pm13$ & $0.55\pm0.03$ &  & 0.71/0.25 & 3.9/0.19 \\
\tnumba & \tname & $0.085\pm0.02$ & $21\pm5$ & $2.71\pm0.05$ & $2.55\pm0.03$ & 0.66/0.092 & 3.6/0.068 \\
\unumba & \uname & $-0.13\pm0.01^{*}$ & $16\pm2$ & $11.03\pm0.06$ & $11.6\pm0.2$ & 4.9/0.082 & 23/0.056 \\
\hline
\end{tabular}
\caption{Summary of long-term variable sources in the \gx\,field. 
Each source, except for \gx\,and \gxpsr, has been given their own MKT name which will be used in this paper, but may change if the source is identified as a known source. 
The names of the sources include the astrometrically corrected (see Section\,\ref{sec: VAR absolute astrometry}) RA and Dec, both the RA and Dec have uncertainties of 0\farcs4.
$\oline{\alpha}$ is the weighted mean MeerKAT spectral index for the source, the sources denoted with a star ($^{*}$) have significantly variable spectral indices. The MeerKAT spectral indices were all produced using simple power-law fits.
The \vardiff\,has been calculated for each source using the ten-epoch binned MFS light curves and the weighted mean.
S$_{\mathrm{909\,MHz}}$ is the mean flux density in the 909\,MHz MeerKAT subband and S$_{\mathrm{RACS}}$ is the RACS flux density with a central frequency of 887.5\,MHz. {We show both the unbinned and binned
$\eta_{\mathrm{MFS}}$ and $V_{\mathrm{MFS}}$ 
values for each source.}
}
\label{tab: VAR supp source spec summary}
\end{table*}

\begin{figure*}
\begin{center}
\includegraphics[width=\textwidth]{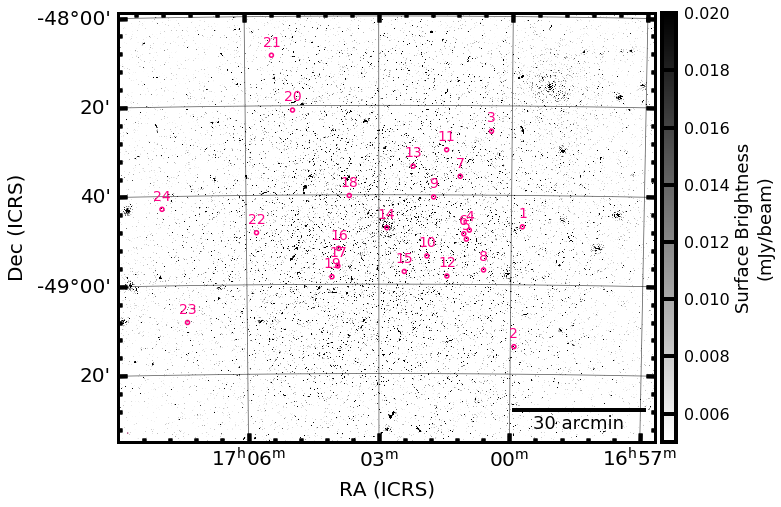}
\caption[MeerKAT deep, MFS image of the \gx\,field showing the positions of the variable sources.]{MeerKAT deep, MFS image of the \gx\,field showing the positions of the variable sources, including the three known variable sources.  The numbers correspond to those in Table\,\ref{tab: VAR supp source spec summary}. The synthesised beam shape is too small to see in this figure and, as this is an MFS image, it has not been primary beam corrected.}
\label{fig: VAR MKT FoV}
\end{center}
\end{figure*}

Three of the long-term variable sources are coincident with a source in at least one of the catalogues shown in Table\,\ref{tab: vizier cats}: \gname\, (source \gnumba), \lname\, (source \lnumba), 
and \rname\, (source \rnumba). Only
\gname\,
is detected by MeerLICHT, and one source, \uname\, (source \unumba), is outside the MeerLICHT FoV. Out of the 24 (including the three known variables) variable sources 11 have matches in RACS, including \gxpsr\,and \gx. The sources and their RACS flux densities are shown in Table\,\ref{tab: VAR supp source spec summary}. The MeerKAT FoV with the locations of the 21 new variable sources, plus the three known variable sources, is shown in Figure\,\ref{fig: VAR MKT FoV}. Note that source \gxnumba\,is \gx, the phase centre of the image. A MeerLICHT image showing its FoV is shown in Figure\,\ref{fig: VAR SUPP MLT image}, where we can see how crowded this field is in the optical. This means that it is particularly interesting that only one of the sources are matched to MeerLICHT sources. {No other sources have clear optical counterparts in the MeerLICHT image}, for postage stamps of the MeerLICHT $q$-band positions of all of the variable sources, see Figures\,\ref{fig: VAR SUPP MLT stamps 1-8} to \ref{fig: VAR SUPP MLT stamps 17-24}. We will now discuss each source individually.

\begin{figure*}
\begin{center}
\includegraphics[width=\textwidth]{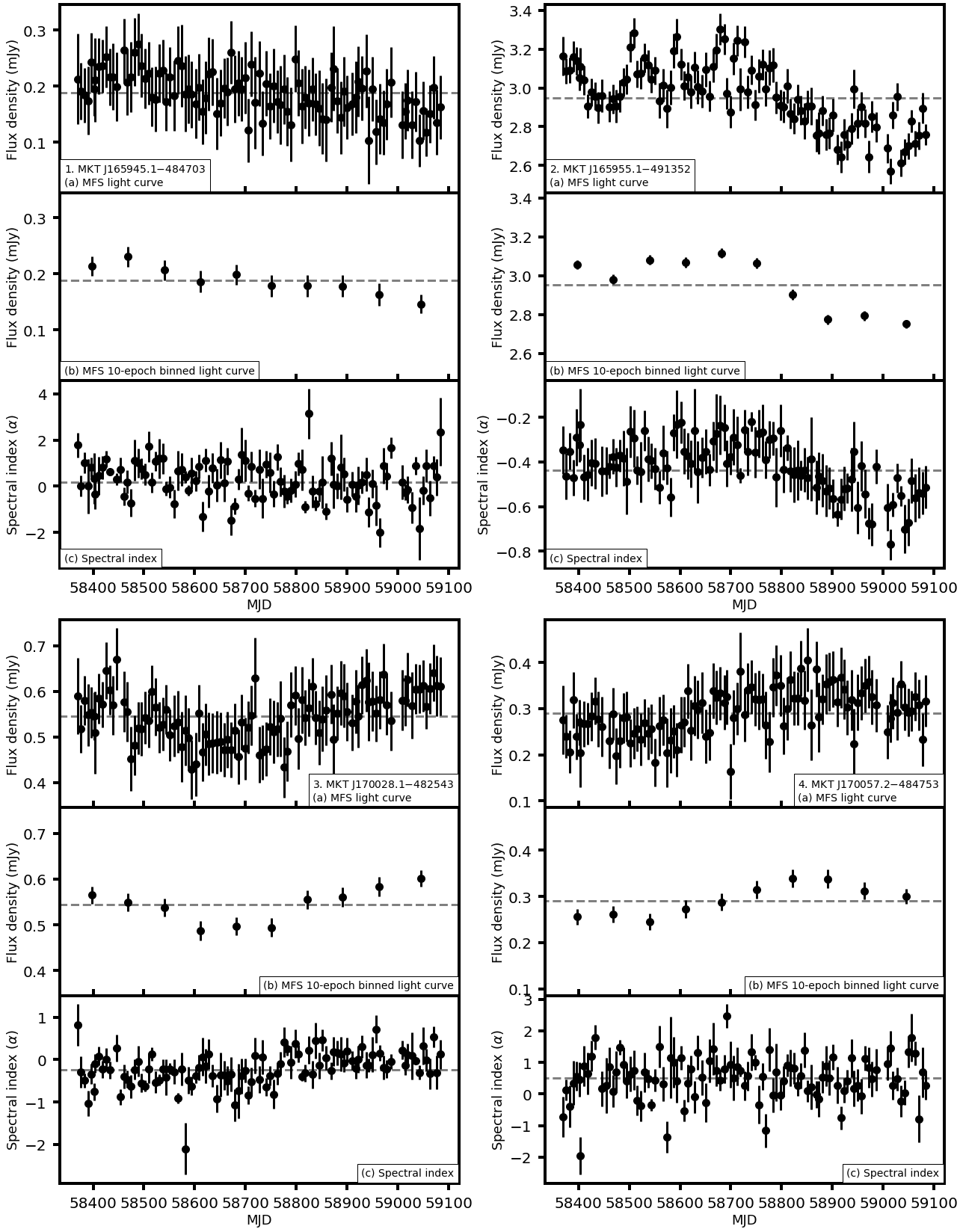}
\caption{Light curves, binned light curves, and spectral index over time for sources:
(\anumba) \aname,
(\bnumba) \bname,
(\cnumba) \cname,
(\dnumba) \dname. The grey-dashed line in each panel shows the weighted mean value.}
\label{fig: 2_3_4_5 large combined}
\end{center}
\end{figure*}

\subsubsection{\aname}
This source (source \anumba) was also manually identified as variable using its MFS light curve, and does not appear variable in the individual subband light curves due to the low S/N. {It has an unbinned/binned $\eta_{\mathrm{MFS}}$ of 0.43/2.0 and $V_{\mathrm{MFS}}$ of 0.2/0.13.} \aname\,has a slowly decreasing flux density over time, see Figure\,\ref{fig: 2_3_4_5 large combined}, and a \vardiff\,of $45\pm13$\,per\,cent. This source has a mean spectral index of  $\oline{\alpha}=0.19\pm0.05$, but $\alpha$ also appears to vary randomly over time due to the low S/N in some subbands. This source does not have any known counterparts at other wavelengths or RACS.

\subsubsection{\bname}
\bname\,(source \bnumba) is an outlier in the \varp\,parameter. {The source has an unbinned/binned $\eta_{\mathrm{MFS}}$ of 4.5/34 and $V_{\mathrm{MFS}}$ of 0.055/0.048.} It has an initially flat light curve that decreases in flux density after approximately a year, shown in Figure\,\ref{fig: 2_3_4_5 large combined}, with a \vardiff\,of $12\pm1$\,per\,cent. There appears to be shorter time scale variability superimposed over the long-term variability; however, this correlates with the light curves of other sources in the field and is therefore likely to be a residual uncorrected correlated effect.
The spectral index varies over time in a way that mirrors the MFS flux density light curve, with a mean value of $\oline{\alpha}=-0.44\pm0.01$.
The RACS flux density is $9.81\pm0.06$\,mJy (with a centre frequency of 887.5\,MHz), which is consistent with the mean MeerKAT flux density that we observe at 909\,MHz: $9.15\pm0.14$\,mJy.
We did not find any counterparts for this source at other wavelengths.

\subsubsection{\cname}
This source (source \cnumba) was found via manual light curve vetting and was picked up at various subbands as well as MFS. {It has an unbinned/binned $\eta_{\mathrm{MFS}}$ of 0.67/3.9 and $V_{\mathrm{MFS}}$ of 0.096/0.072.} The light curve of this source initially decreases and then slowly increases over time with a \vardiff\,of $21\pm5$\,per\,cent, see Figure\,\ref{fig: 2_3_4_5 large combined}.
The mean spectral index is $\oline{\alpha}=-0.24\pm0.02$.
Due to the low S/N in some subbands there is scatter in the spectral index over time, including an outlier at MJD$\sim$56000 which was caused by a noisy epoch in the 1016\,MHz subband, but has no overall variability in the spectral index. 
There are no multi-wavelength counterparts.

\subsubsection{\dname}
This source (source \dnumba) was found manually in the MFS band. {It has an unbinned/binned $\eta_{\mathrm{MFS}}$ of 0.69/3.3 and $V_{\mathrm{MFS}}$ of 0.17/0.11.} The light curve of this source, shown in Figure\,\ref{fig: 2_3_4_5 large combined}, increases from approximately MJD 58550 to 58700 before roughly levelling off again and has a \vardiff\,of $32\pm9$\,per\,cent. \dname\,has a mean spectral index of $\oline{\alpha}=0.50\pm0.05$, with scatter on the spectral index over time caused by the low S/N in some subbands.
This source does not have any multi-wavelength or RACS counterparts.


\begin{figure*}
\begin{center}
\includegraphics[width=\textwidth]{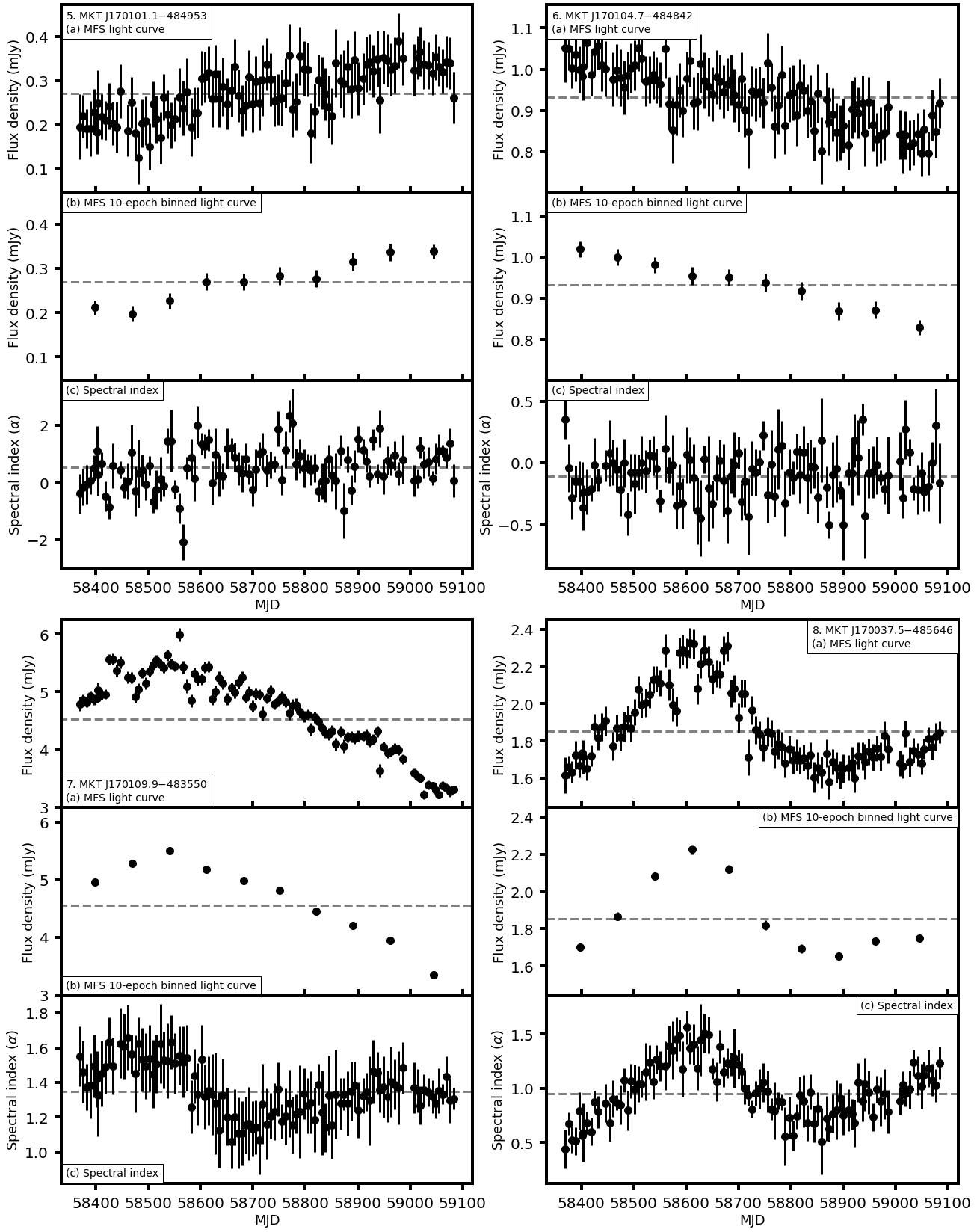}
\caption{Light curves, binned light curves, and spectral index over time for sources:
(\enumba) \ename,
(\fnumba) \fname,
(\gnumba) \gname,
(\hnumba) \hname.
The grey-dashed line in each panel shows the weighted mean value.}
\label{fig: 6_7_8_9 large combined}
\end{center}
\end{figure*}

\subsubsection{\ename}
\ename\,(source \enumba) was identified manually in the MFS light curves. {It has an unbinned/binned $\eta_{\mathrm{MFS}}$ of 1.0/7.7 and $V_{\mathrm{MFS}}$ of 0.21/0.18.} The sources rises slowly over time, see Figure\,\ref{fig: 6_7_8_9 large combined}, and has a \vardiff\,of $52\pm9$\,per\,cent.
\dname\,has a mean spectral index of $\oline{\alpha}=0.54\pm0.05$, with scatter on the spectral index over time caused by the low S/N in some subbands.
We do not find any multi-wavelength counterparts for this source and it is not detected by RACS.

\subsubsection{\fname}
This source (source \fnumba) was identified as variable as it is an outlier in the \varp\,parameter. {It has an unbinned/binned $\eta_{\mathrm{MFS}}$ of 1.3/11 and $V_{\mathrm{MFS}}$ of 0.074/0.065.} The flux density decreases over time with a \vardiff\,of $20\pm3$\,per\,cent, the light curve is shown in Figure\,\ref{fig: 6_7_8_9 large combined}. The spectral index is scattered around a mean of $\oline{\alpha}=-0.11\pm0.02$. This source has a MeerKAT 909\,MHz mean flux density of $1.12\pm0.19$\,mJy and was detected by RACS with an 887.5\,MHz flux density of $1.96\pm0.06$\,mJy.
We did not find any coincident sources at other wavelengths.

\subsubsection{\gname}
\begin{figure}
\begin{center}
\includegraphics[width=\columnwidth]{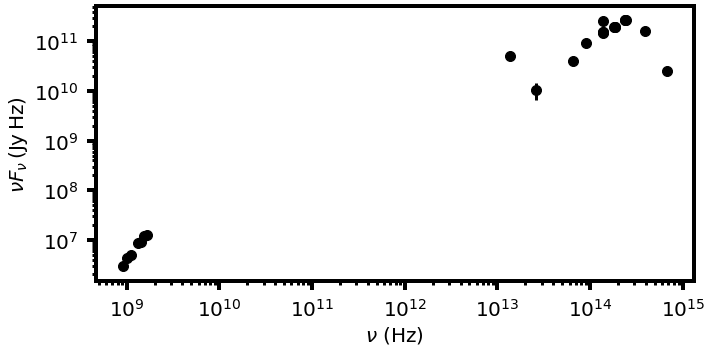}
\caption{Spectral energy distribution of \gname\,(source \gnumba) including the archival MeerLICHT and
MeerKAT observations. 
The MeerKAT flux densities shown here are the mean flux densities 
over every epoch these sources are observed. 
\gname\,is only detected in the MeerLICHT $i$-band,
the non-detections are not shown here. 
The MeerLICHT optical flux density has been extinction-corrected using the 
NASA/IPAC Extragalactic Database Coordinate 
Transformation and Galactic Extinction Calculator.}
\label{fig: VAR MKTJ170109.9-483550 MKTJ170222.5-483144 SED}
\end{center}
\end{figure}

\gname\,(source \gnumba) is highly variable in all MeerKAT subbands, shown in Figure\,\ref{fig: 6_7_8_9 large combined}, and is an outlier in the \varp\,parameter. {It has an unbinned/binned $\eta_{\mathrm{MFS}}$ of 59/620 and $V_{\mathrm{MFS}}$ of 0.14/0.14.} The flux density of the source increases over the first $\sim$100 days of MeerKAT observations and then decreases, it has a \vardiff\,of $47\pm1$\,per\,cent. 
The spectral index varies between $\alpha\sim1$ and $\alpha\sim1.8$, with a different trend to the flux density variations and a mean of $\oline{\alpha}=1.35\pm0.01$.
This source has a MeerKAT 909\,MHz mean flux density of $3.25\pm0.18$\,mJy and was detected by RACS with an 887.5\,MHz flux density of $3.32\pm0.05$\,mJy.
This source is coincident with a 2MASS source, 2MASS 17010986-4835508, with a separation of 0\farcs39.
2MASS 17010986-4835508 has a K-magnitude of 14.465$\pm$0.123, but has poor-quality photometry in the J and H bands \citep{2003yCat.2246....0C}. The 2MASS source was matched to an AllWISE source, AllWISE\,J170109.82-483550.8, by the TESS Input Catalogue \citep[TIC;][]{2019AJ....158..138S} with W1 magnitude 13.670$\pm$0.040 and W2 magnitude 13.617$\pm$0.046. The 2MASS/AllWISE source is classified as a star by the TIC due to its point-like morphology. \gname\,is detected in the MeerLICHT $i$-band image with an AB\,magnitude of 18.90$\pm$0.04. 
The spectral energy distribution (SED) including the archival, MeerLICHT, and MeerKAT data for \gname\,is shown in Figure\,\ref{fig: VAR MKTJ170109.9-483550 MKTJ170222.5-483144 SED}.


\subsubsection{\hname}
\hname\,(source \hnumba) was found as it is an outlier in the \varp\,parameter in the 1337 to 1658\,MHz subbands and MFS. {It has an unbinned/binned $\eta_{\mathrm{MFS}}$ of 7.5/73 and $V_{\mathrm{MFS}}$ of 0.11/0.11.} The flux density of the source increases sharply until approximately MJD 58600, decreases sharply until approximately MJD 58800, and then slowly increases. \hname\,has a \vardiff\,of $31\pm2$\,per\,cent and the light curve is shown in Figure\,\ref{fig: 6_7_8_9 large combined}.
The spectral index varies between $\alpha\sim0.5$ and $\alpha\sim1.5$ with a similar shape to the light curve and a mean of $\oline{\alpha}=0.95\pm0.01$.
We do not find any multi-wavelength counterparts for this source.


\subsubsection{\iname}
\begin{figure*}
\begin{center}
\includegraphics[width=\textwidth]{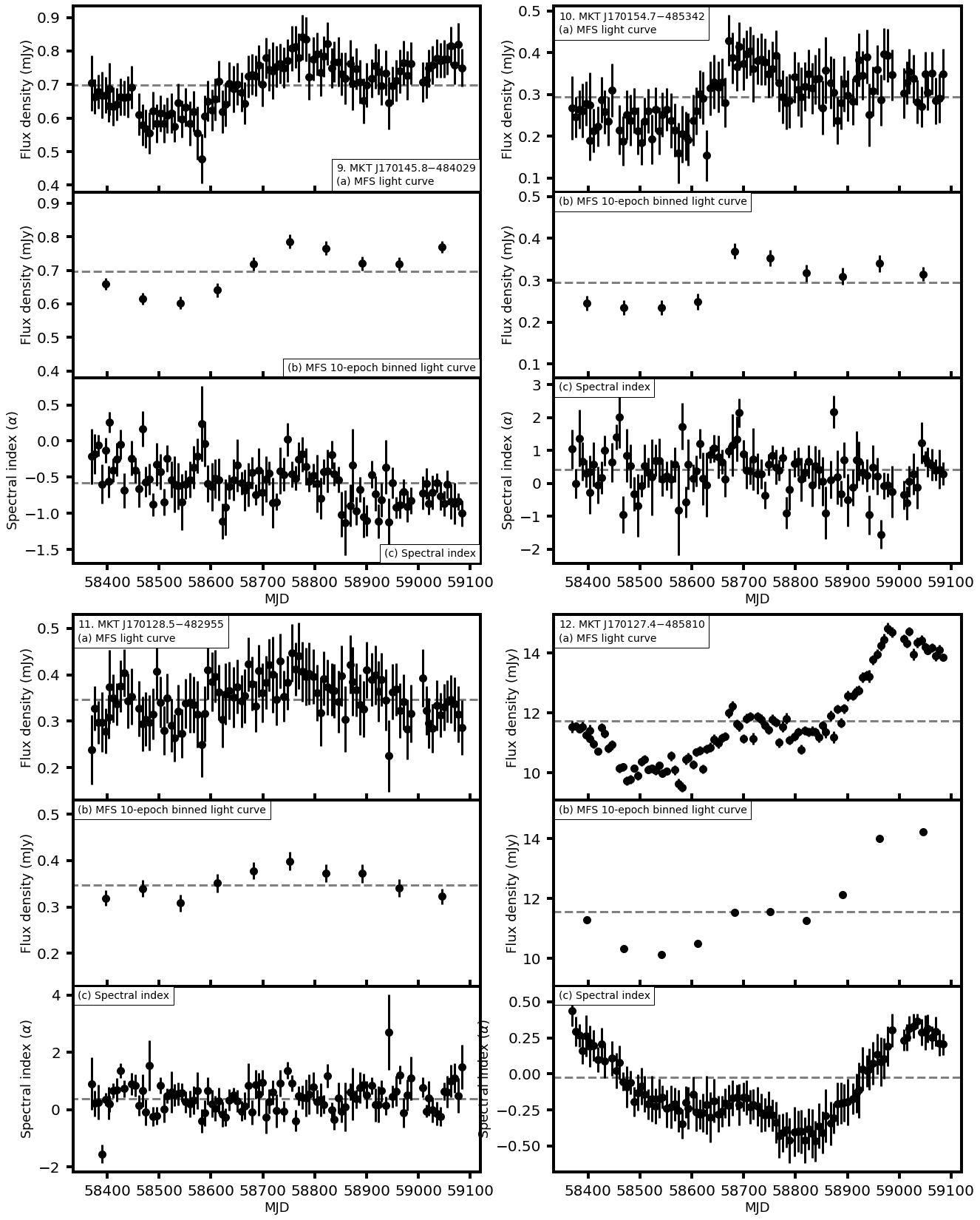}
\caption{Light curves, binned light curves, and spectral index over time for sources: 
(\inumba) \iname,
(\jnumba) \jname,
(\knumba) \kname,
(\lnumba) \lname.
The grey-dashed line in each panel shows the weighted mean value.}
\label{fig: 10_11_12_13 large combined}
\end{center}
\end{figure*}

This source (source \inumba) was found through manual vetting, but is a slight outlier in the unbinned \varp\,parameter in the 1016\,MHz band and MFS. {It has an unbinned/binned $\eta_{\mathrm{MFS}}$ of 1.4/12 and $V_{\mathrm{MFS}}$ of 0.1/0.095.} 
The flux density of \iname\,decreases for approximately 100\,days, increases for approximately 300\,days, decreases for approximately 100\,days, followed by a further increase, see Figure\,\ref{fig: 10_11_12_13 large combined}. The \vardiff\,is $26\pm4$\,per\,cent.
The spectral index roughly decreases for two years with a mean spectral index of $\oline{\alpha}=-0.58\pm0.02$.
We do not find any multi-wavelength counterparts for this source.

\subsubsection{\jname}
This source (source \jnumba) was found via manual vetting. {It has an unbinned/binned $\eta_{\mathrm{MFS}}$ of 1.1/7.7 and $V_{\mathrm{MFS}}$ of 0.21/0.17.} \jname\,has a constant flux density until approximately MJD 56800 at which point the flux density increases for approximately 100 days before decreasing slightly again and returning to roughly constant. The light curve is shown in Figure\,\ref{fig: 10_11_12_13 large combined} and it has a \vardiff\,of $46\pm9$\,per\,cent.
The source has a mean spectral index of $\oline{\alpha}=0.43\pm0.05$, with some scatter caused by the low S/N in some subbands.
This source is not detected by RACS and does not have any multi-wavelength counterparts.

\subsubsection{\kname}
\kname\,(source \knumba) was found through manual vetting. {It has an unbinned/binned $\eta_{\mathrm{MFS}}$ of 0.57/2.6 and $V_{\mathrm{MFS}}$ of 0.13/0.084.} The \vardiff\,of the light curve is $26\pm8$\,per\,cent.
The light curve, shown in Figure\,\ref{fig: 10_11_12_13 large combined}, increases slightly for approximately a year, before slowly decreasing again.
The mean spectral index is $\oline{\alpha}=0.38\pm0.04$, which is constant over time apart from scatter due to the low subband S/N.
This source is not detected in RACS or MeerLICHT and does not have any archival multi-wavelength counterparts.

\subsubsection{\lname}
\lname\,(source \lnumba) was found as it is an outlier in the \varp\,parameter in all subbands and MFS.
{It has an unbinned/binned $\eta_{\mathrm{MFS}}$ of 79/805 and $V_{\mathrm{MFS}}$ of 0.12/0.12.}
The MFS light curve initially decreases before increasing over approximately a year, and begins decreasing again in the last two months, see Figure\,\ref{fig: 10_11_12_13 large combined}. The \vardiff\,of the light curve is $35\pm1$\,per\,cent.
The spectral index varies between $\alpha\sim-0.5$ and $\alpha\sim0.5$ with a mean of $\oline{\alpha}=-0.02\pm0.01$. There is some curvature in the spectrum of this source, and as such the power law spectral index fit is an approximation. The spectral index decreases over about 400 days before increasing again, followed by a decrease echoing the flux density decrease. 
\lname\,is detected by RACS, with a flux density of $14.32\pm0.6$\,mJy at 887.5\,MHz. The mean MeerKAT flux density of the source at 909\,MHz is $13.72\pm0.14$, which is roughly consistent with the RACS flux density.
This source is coincident with unWISE sources 2542m485o0077626 and unWISE source 2564m485o0076399 with a separation of 0\farcs2 for both. \lname\,is not coincident with any other sources in Vizier and is not detected by MeerLICHT.


\subsubsection{\mname}
\begin{figure*}
\begin{center}
\includegraphics[width=\textwidth]{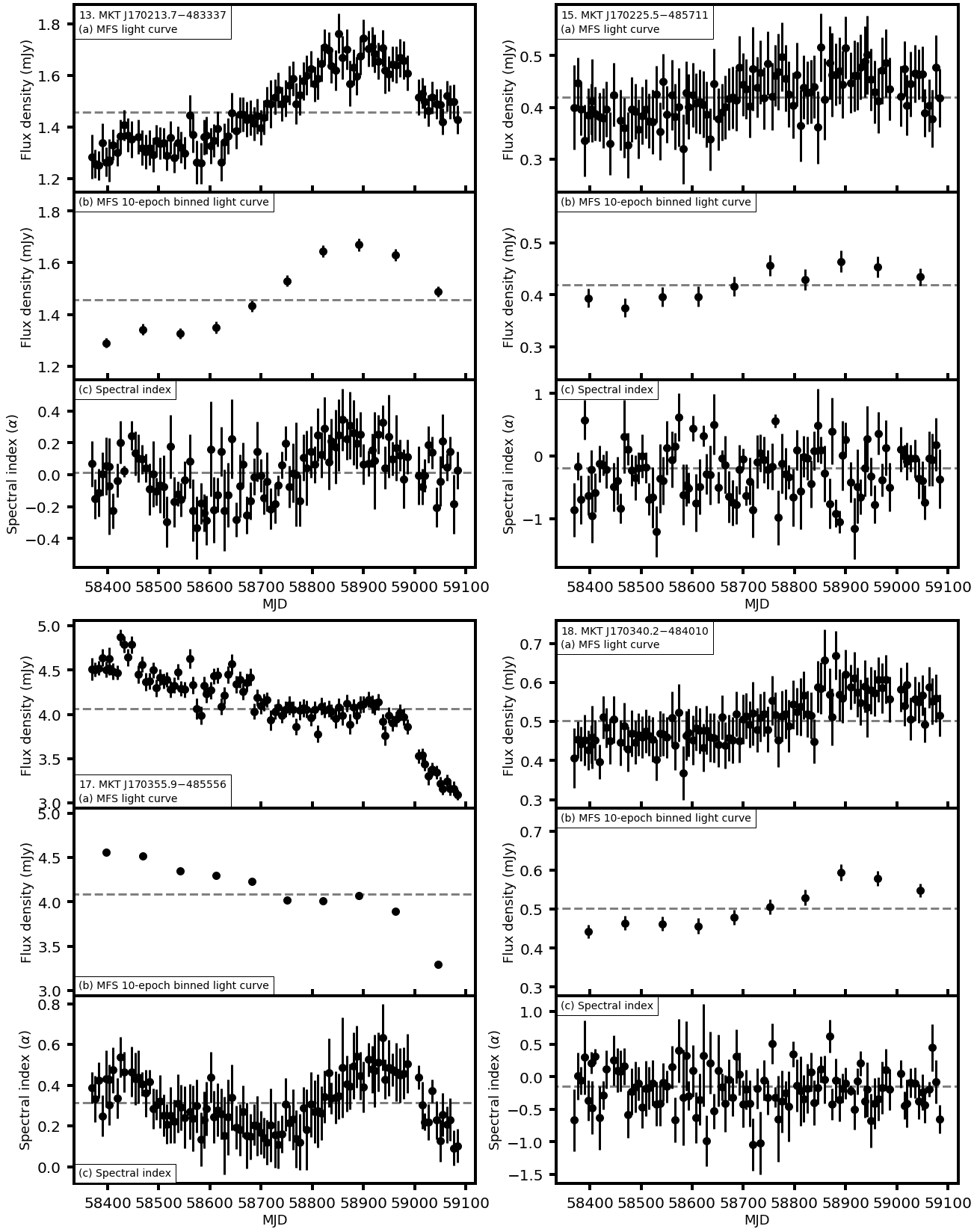}
\caption{Light curves, binned light curves, and spectral index over time for sources: 
(\mnumba) \mname,
(\nnumba) \nname,
(\onumba) \oname,
(\pnumba) \pname,
The grey-dashed line in each panel indicates the weighted mean value.}
\label{fig: 14_18_21_22 large combined}
\end{center}
\end{figure*}

\mname\,(source \mnumba) was found as it is a clear outlier in the \varp\,parameter in the 1658\,MHz subband and MFS.
{It has an unbinned/binned $\eta_{\mathrm{MFS}}$ of 4.3/43 and $V_{\mathrm{MFS}}$ of 0.097/0.098.}
Its light curve, shown in Figure\,\ref{fig: 14_18_21_22 large combined}, is roughly constant for the first $\sim$200\,days of MeerKAT observations before increasing for approximately a year, followed by a decrease until the end of the MeerKAT light curve. It has a \vardiff\,of $26\pm2$\,per\,cent.
The spectral index varies between $\alpha\sim-0.5$ and $\alpha\sim0.5$ with a mean spectral index of $\oline{\alpha}=0.01\pm0.01$. The evolution of the spectral index roughly mirrors that of the light curve.
The source does not have any multi-wavelength counterparts and is not detected by MeerLICHT or RACS.

\subsubsection{\nname}

The source (source \nnumba) was found via manual vetting.
{It has an unbinned/binned $\eta_{\mathrm{MFS}}$ of 0.5/2.5 and $V_{\mathrm{MFS}}$ of 0.1/0.073.}
The flux density of \nname\,increases slowly over time, see Figure\,\ref{fig: 14_18_21_22 large combined}, with a \vardiff\,of $21\pm7$\,per\,cent.
The mean spectral index of the source is $\oline{\alpha}=-0.19\pm0.03$ with scatter caused by low S/N in some subbands.
This source was not detected by RACS or MeerLICHT, and no multi-wavelength counterparts were found.




\subsubsection{\oname}

This source (source \onumba) is an outlier in the \varp\,parameter in all subbands and MFS. 
{It has an unbinned/binned $\eta_{\mathrm{MFS}}$ of 20/200 and $V_{\mathrm{MFS}}$ of 0.09/0.089.}
The source's light curve, shown in Figure\,\ref{fig: 14_18_21_22 large combined}, decreases over two years.
The \vardiff\,of \oname\,is $31\pm1$\,per\,cent.
This source has a mean spectral index of $\oline{\alpha}=0.31\pm0.01$, and the spectral index varies between $\sim0$ and $\sim0.7$.
There is some curvature in the spectrum of this source, and as such the power law spectral index fit is an approximation.
The mean MeerKAT 909\,MHz flux density is $4.01\pm0.18$\,mJy and the source is detected by RACS with an 887.5\,MHz flux density of $4.73\pm0.06$\,mJy.
\oname\,is not detected by MeerLICHT and we do not find any counterparts using Vizier.

\subsubsection{\pname}

We found \pname\,(source \pnumba) using manual vetting.
{It has an unbinned/binned $\eta_{\mathrm{MFS}}$ of 1.0/8.1 and $V_{\mathrm{MFS}}$ of 0.12/0.11.}
The MFS light curve of the source, shown in Figure\,\ref{fig: 14_18_21_22 large combined}, is approximately constant for 300\,days, increases for approximately 200\,days, then decreases slightly until the end of the MeerKAT light curve. The \vardiff\,of the light curve is $30\pm5$\,per\,cent.
The mean spectral index of the source is $\oline{\alpha}=-0.14\pm0.03$ with scatter caused by low S/N in some subbands.
We do not find any Vizier, RACS, or MeerLICHT counterparts for this source.

\subsubsection{\qname}

\begin{figure*}
\begin{center}
\includegraphics[width=\textwidth]{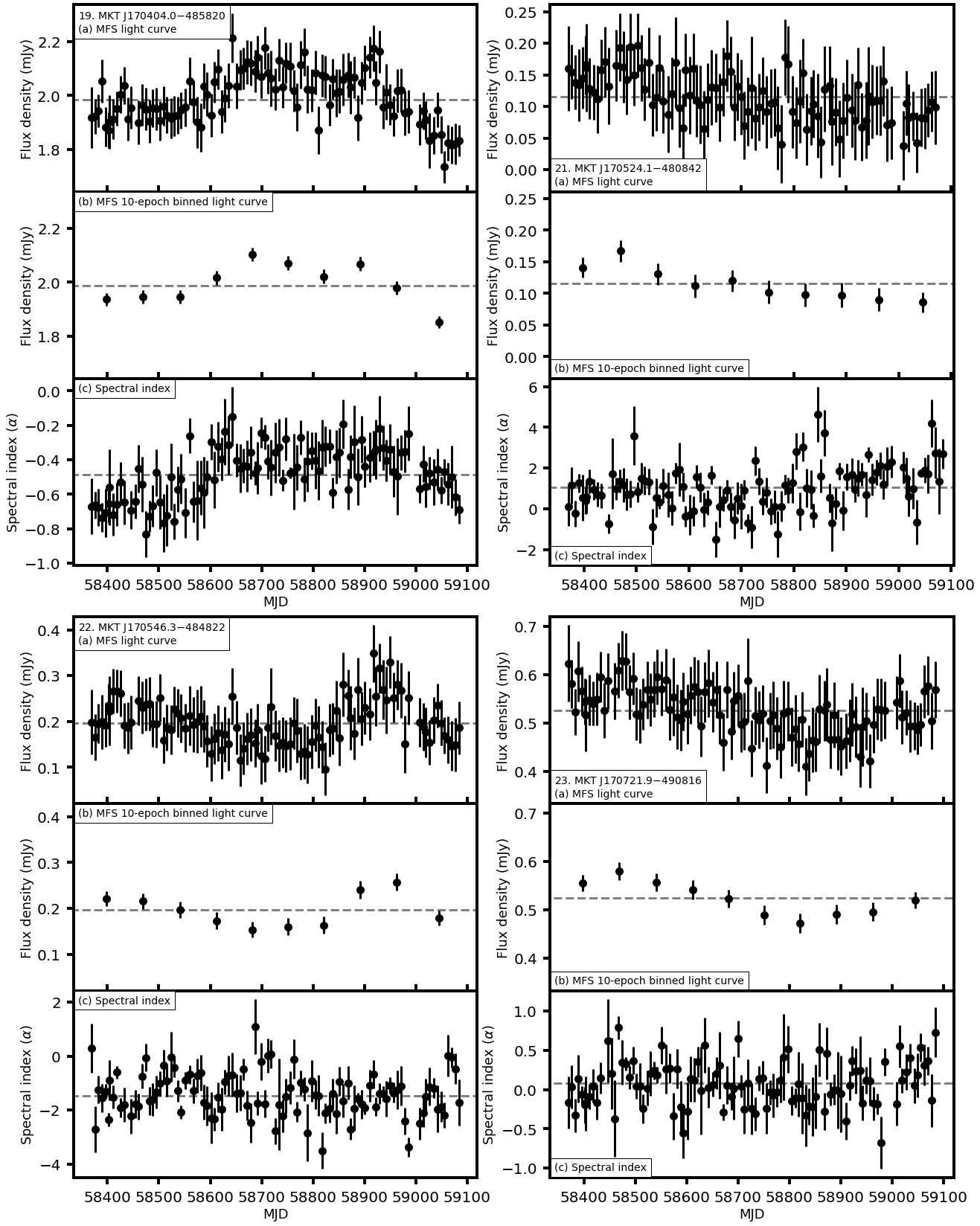}
\caption{Light curves, binned light curves, and spectral index over time for sources:
(\qnumba) \qname,
(\rnumba) \rname, 
(\snumba) \sname, 
(\tnumba) \tname. The grey-dashed line in each panel shows the weighted mean value.}
\label{fig: 23_25_26_27 large combined}
\end{center}
\end{figure*}

\qname\,(source \qnumba) was found via manual vetting.
{It has an unbinned/binned $\eta_{\mathrm{MFS}}$ of 1.5/10 and $V_{\mathrm{MFS}}$ of 0.047/0.038.}
The source is constant for approximately 200\,days, increases for approximately 100\,days, remains constant for approximately 300\,days, then decreases. The light curve is shown in Figure\,\ref{fig: 23_25_26_27 large combined} with a \vardiff\,of $13\pm2$\,per\,cent.
The spectral index varies in a similar way to the flux density, varying between $\alpha\sim-1$ and $\alpha\sim0$ with a mean of $\oline{\alpha}=-0.49\pm0.01$.
The mean MeerKAT flux density at 909\,MHz is $2.81\pm0.18$\,mJy and the RACS 887.5\,MHz flux density is $2.96\pm0.03$\,mJy.
We do not find any Vizier or MeerLICHT counterparts.

\subsubsection{\rname}

\rname\,(source \rnumba) was found using manual vetting and has a decreasing flux density over time.
{It has an unbinned/binned $\eta_{\mathrm{MFS}}$ of 0.46/2.3 and $V_{\mathrm{MFS}}$ of 0.32/0.22.}
The light curve of the source is shown in Figure\,\ref{fig: 23_25_26_27 large combined}.
The \vardiff\,of the light curve is $70\pm20$\,per\,cent.
The mean spectral index of the source is $\oline{\alpha}=1.06\pm0.06$ with scatter caused by low S/N in some subbands.
This source is coincident with unWISE source 2564m485o0191898 (0\farcs138 separation). We do not find any other Vizier, MeerLICHT, or RACS counterpart for this source.

\subsubsection{\sname}

\sname\,(source \snumba) was found using manual vetting.
{It has an unbinned/binned $\eta_{\mathrm{MFS}}$ of 0.71/3.9 and $V_{\mathrm{MFS}}$ of 0.25/0.19.}
The light curve, shown in Figure\,\ref{fig: 23_25_26_27 large combined}, slowly decreases over the first half of the light curve before increasing and decreasing more steeply in the second half. The \vardiff\,of the light curve is $53\pm13$\,per\,cent.
The mean spectral index of the source is $\oline{\alpha}=-1.46\pm0.05$ with scatter caused by low S/N in some subbands.
This source is not detected by MeerLICHT or in archival Vizier catalogues and is not detected by RACS.

\subsubsection{\tname}

This source (source \rnumba) was found using manual vetting.
{It has an unbinned/binned $\eta_{\mathrm{MFS}}$ of 0.66/3.6 and $V_{\mathrm{MFS}}$ of 0.092/0.068.}
The light curve of \tname, shown in Figure\,\ref{fig: 23_25_26_27 large combined}, decreases over time before slowly increasing for the last $\sim$200\,days.
The mean spectral index of the source is $\oline{\alpha}=0.08\pm0.02$.
There is some curvature in the spectrum of this source, particularly towards the top of the MeerKAT band, and as such the power law spectral index fit is an approximation.
The mean 909\,MHz MeerKAT flux density is $2.69\pm0.14$\,mJy and the RACS 887.5\,MHz flux density is $2.55\pm0.03$\,mJy.
This source does not have a Vizier or MeerLICHT counterpart.

\subsubsection{\uname}

\begin{figure}
\begin{center}
\includegraphics[width=\columnwidth]{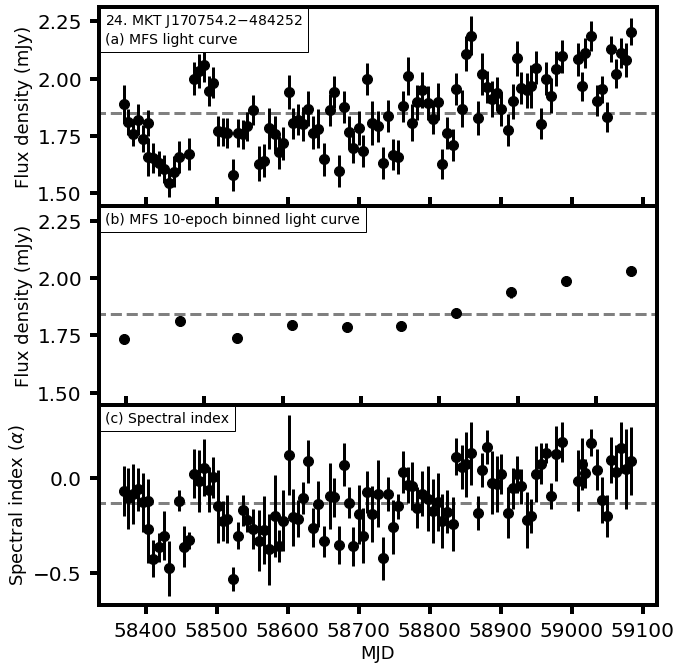}
\caption{Light curve, binned light curve, and spectral index over time for source (\unumba) \uname.
The grey-dashed line in each panel shows the weighted mean value.}
 \label{fig: 28 large combined}
 \end{center}
 \end{figure}

\uname\,(source \unumba) was found via manual vetting.
{It has an unbinned/binned $\eta_{\mathrm{MFS}}$ of 4.9/23 and $V_{\mathrm{MFS}}$ of 0.082/0.056.}
The source has an increasing flux density, shown in Figure\,\ref{fig: 28 large combined}.
The light curve of this source also shows shorter time scale variability; however, we note that this variability correlates with some other light curves in the field, indicating that it is caused by residual systematics and is not intrinsic to the source.
The spectral index increases over time from $\alpha\sim-0.5$ to $\alpha\sim0.5$ with a mean of $\oline{\alpha}=-0.13\pm0.01$.
There is some curvature in the spectrum of this source, particularly towards the top of the MeerKAT band, and as such the power law spectral index fit is an approximation.
\uname\,has a mean MeerKAT 909\,MHz flux density of $11.04\pm0.12$\,mJy and is detected by RACS with a flux density of $11.60\pm0.15$\,mJy.
The source falls outside the MeerLICHT FoV and we did not find any multi-wavelength counterparts in Vizier.

\section{Discussion}
\label{sec: VAR discussion}

{In Section\,\ref{sec: VAR sys effects} we discussed the discovery and mitigation of systematic effects in the light curves of sources in the \gx\,field (see Appendix\,\ref{app: systematics} for further information). While we were investigating these systematics we manually looked through plots of all of the unbinned light curves of sources with S/N>3 in at least one epoch, with the mean value also plotted to enable easier identification of systematics and variability. In doing this, we first identified many of the long-term variable sources presented in this paper. This is why many of the sources in Section\,\ref{sec: VAR results} were found via manual vetting. We performed the binning (Section\,\ref{sec: binning method}) after many of the sources had already been identified manually as candidate long-term variable sources, but all 21 of the new long-term variable sources have binned \varp\,values $>1$, and as such would have been identified as candidates for further investigation using the binned variability parameters. The \varp\,value of all of the new variable sources increased after binning, which indicates that there is a long-term trend in the light curves. This is because binning enhances the \varp\,value where there is a trend on time scales comparable to the binning.}

The non-variable sources in Figure\,\ref{fig: VAR var params} are clustered around \varp$\sim0.3$ instead of the expected \varp$\sim1.0$ for the reduced $\chi^{2}$. 
{We believe that this is because the uncertainties on the light curves and systematics are conservative, and when we propagate these it leads to more conservative binned light curve uncertainties.
This means that the \varp\,values of the variable sources are also likely to be under-estimated, which is why we consider the sources in comparison with each other.}

If we include the known variable sources \gx, \gxpsr, and \fb\,we find that {$2.2\pm0.5$\,per\,cent} of the point sources in the field with S/N$>3$ are long-term variable sources.
Surveys such as \citet{VLA_Carilli}, \citet{1994ApJ...437L..43F} and \citet{2013ApJ...768..165M} found that 1--2\,per\,cent of radio sources in their surveys were variable. 
{This is consistent with the percentage of sources we find; however, we have only included known variables and long-term variables in this work and we have three more years of weekly monitoring of this field to search for variable sources. This implies that $2.2\pm0.5$\,per\,cent is the lower limit on the percentage of variable sources in the field.}

There are over $400000$ optical sources in the deep MeerLICHT image, which gives $\sim40$ sources per square arcminute or 0.01 sources per square arcsecond. {Initially it may be surprising that we did not find more matches with such a high density of optical sources. There are $\sim0.3$ MeerKAT sources per square arcminute or $\sim8\times10^{-5}$ sources per square arcsecond within a degree of the phase centre. The 0\farcs4 uncertainty on the MeerKAT astronometry gives a $1\sigma$ region of 0.5 square arcseconds around each radio source. We expect 0.006 MeerLICHT sources per 0.5 square arcseconds. We find 1080 unique MeerKAT sources, if we multiply this by 0.006 we only expect 6.5 random matches between MeerKAT and MeerLICHT sources. Similarly, we would expect 0.14 matches between the 24 variable radio sources in the FoV and MeerLICHT sources. Instead, we find that three of the variable sources (including \gx\,and \fb) in the field match with MeerLICHT sources, which is 20 times the expected 0.14 matches. We can see the positions of the MeerKAT sources in the} MeerLICHT $q$-band postage stamps of each source in Figures\,\ref{fig: VAR SUPP MLT stamps 1-8} to \ref{fig: VAR SUPP MLT stamps 17-24}. We are looking through the Galactic plane, which means that optical sources outside the Galaxy or on the other side of the Galaxy will be strongly affected by extinction. A lack of optical detections for 20 of the new variable sources does not conclusively mean that the sources are extra-galactic, but this could be why we do not find optical counterparts despite the deep, sensitive MeerLICHT observations.

The ASKAP intra-hour variability study by \citet{2021arXiv210106048W} found correlated variability in sources in a linear formation on the sky, leading them to conclude that the variability is due to structure in the ISM. While some sources, such as \nname\,(source \nnumba), and \pname\,(source \pnumba) have similar variability trends, they are not in the same region on the sky and other variable sources with different trends are between these sources on the sky. We therefore do not find evidence of an ISM scintillation structure in this field based on long-term variability trends.

\citet{VLA_Thyagarajan} found that $\sim1409$ of the 1627 ($\sim87$\,per\,cent) variable sources they found using the FIRST survey were either confirmed to be or consistent with galaxies and QSOs, indicating that the majority of radio variable sources are likely to be AGN. 
The SED of \gname\,(shown in Figure\,\ref{fig: VAR MKTJ170109.9-483550 MKTJ170222.5-483144 SED}) is consistent with a radio-quiet AGN or star-forming galaxy, despite being classified as a star by the TIC based on the round morphology. 
The 20 other new variable sources do not have enough information to confirm their nature; however, we can compare the radio variability to the blazar variability found by the \textit{Fermi}-GST AGN Multi-frequency Monitoring Alliance \citep[F-GAMMA;][]{2016A&A...596A..45F}. F-GAMMA has observed $\sim60$ selected blazars (plus targets of opportunity) approximately monthly from 2007 to 2015 in radio frequency bands from 2.64\,GHz to 43\,GHz.
They find that the radio variability of blazars is due to shocks in the relativistic jet at high frequencies and refractive scintillation at low frequencies \citep{2012JPhCS.372a2007A}. We can see in their released light curves that the variability of the blazars they observe can be extreme at the higher (43\,GHz) frequencies, but has a lower \vardiff\,and a smoother, slower variability at low frequencies (2.64\,GHz). We note that the shape and time scales of variability in their sources at low frequencies is similar to some of our sources, and that the spectral index evolution in some sources also mirrors the variability of our sources. While we cannot yet determine the nature of our sources or the cause of the variability, the many similarities between our sources and the F-GAMMA sources means that refractive scintillation of AGN could explain what we observe. 

{We used the model\footnote{\href{https://github.com/PaulHancock/RISS19}{https://github.com/PaulHancock/RISS19}} by \citet{2019arXiv190708395H} to predict the approximate variability time scale and $V_{\mathrm{MFS}}$ for scintillating extra-galactic sources at the position of each of our sources. \citet{2019arXiv190708395H} use 
{an H$_{\alpha}$ intensity map} \citep{2003ApJS..146..407F} to predict the effect of RISS on radio sources in searches for variable sources. They find that more RISS-induced variable sources are expected for searches close to the Galactic plane than further away. Table\,\ref{tab: RISS predications} shows the results of using their code to predict the RISS variability time scale and expected $V_{\mathrm{MFS}}$ after a year of observations at the positions of each of our sources. The predicted $V_{\mathrm{MFS}}$ is the maximum expected value, and multiple sources can have the same predicted values due to the coarseness of the model grid.  The known Galactic sources, \gx, \gxpsr\,and \fb, are intrinsically variable and are not variable due to RISS. As expected, the predicted RISS \modp\,values and timescales for these sources do not match the observed \modp\,values and timescales.
The predicted RISS \modp\,values are consistent with the observed, binned \modp\,values for 18 of the 21 new variable sources.
{Some sources have a lower than expected \modp\,compared to the RISS values, such as sources \bname\, (source \bnumba) and \lname\, (source \lnumba). This could be because those sources are Galactic sources, or it could be because the true electron density along the line of sight of the source is not well-modelled by {the H$_{\alpha}$ intensity map}.}
\rname\,(source \rnumba) is the only new variable source where the observed \modp\,(0.22 for the binned light curve) is larger than the predicted \modp\,($0.19\pm0.005$).
The predicted RISS timescales for all 25 of the new variable sources are on the order of a year, which matches the observed timescales.}

\begin{table*}
\centering
\begin{tabular}{llrrrr}
 & Name & $V_{\mathrm{MFS}}$ (unbinned) & $V_{\mathrm{MFS}}$ (binned) &$V_{\mathrm{MFS,max}}$ (predicted per year) & Predicted time scale (year)  \\
\hline
\hline
 \anumba & \aname & 0.2 & 0.13 & $0.13\pm0.01$ & $1.3\pm0.1$ \\
 \bnumba & \bname & 0.055 & 0.048 & $0.17\pm0.01$ & $1.1\pm0.08$ \\
\cnumba & \cname & 0.096 & 0.072 & $0.12\pm0.009$ & $1.4\pm0.1$ \\
\dnumba & \dname & 0.17 & 0.11 & $0.18\pm0.01$ & $1.0\pm0.08$ \\
\enumba & \ename & 0.21 & 0.18 & $0.19\pm0.005$ & $0.95\pm0.08$ \\
\fnumba & \fname & 0.074 & 0.065 & $0.19\pm0.005$ & $0.95\pm0.08$ \\
\gnumba & \gname & 0.14 & 0.14 & $0.16\pm0.01$ & $1.1\pm0.09$ \\
\hnumba & \hname & 0.11 & 0.11 & $0.17\pm0.01$ & $1.1\pm0.08$ \\
\inumba & \iname & 0.1 & 0.095 & $0.19\pm0.005$ & $0.92\pm0.08$ \\
\jnumba & \jname & 0.21 & 0.17 & $0.19\pm0.005$ & $0.85\pm0.07$ \\
\knumba & \kname & 0.13 & 0.084 & $0.16\pm0.01$ & $1.1\pm0.09$ \\
\lnumba & \lname & 0.12 & 0.12 & $0.19\pm0.005$ & $0.89\pm0.07$ \\
\mnumba & \mname & 0.097 & 0.098 & $0.19\pm0.005$ & $0.9\pm0.07$ \\
\gxnumba & \gx & 1.9 & 1.4 & $0.2\pm0.006$ & $0.8\pm0.07$ \\
\nnumba & \nname & 0.1 & 0.073 & $0.2\pm0.005$ & $0.83\pm0.07$ \\
\gxpsrnumba & \gxpsr & 0.34 & 0.15 & $0.2\pm0.006$ & $0.77\pm0.07$ \\
\onumba & \oname & 0.09 & 0.089 & $0.2\pm0.006$ & $0.75\pm0.06$ \\
\pnumba & \pname & 0.12 & 0.11 & $0.2\pm0.006$ & $0.8\pm0.07$ \\
\qnumba & \qname & 0.047 & 0.038 & $0.2\pm0.006$ & $0.75\pm0.06$ \\
\fbnumba & \fb & 0.65 & 0.31 & $0.19\pm0.005$ & $0.89\pm0.07$ \\
\rnumba & \rname & 0.32 & 0.22 & $0.19\pm0.005$ & $0.84\pm0.07$ \\
\snumba & \sname & 0.25 & 0.19 & $0.2\pm0.005$ & $0.78\pm0.06$ \\
\tnumba & \tname & 0.092 & 0.068 & $0.2\pm0.006$ & $0.69\pm0.06$ \\
\unumba & \uname & 0.082 & 0.056 & $0.2\pm0.005$ & $0.76\pm0.06$ \\
\hline
    \end{tabular}
    \caption{{Predicted $V_{\mathrm{MFS,max}}$ and variability time scales at 1284\,MHz caused by RISS using the \citet{2019arXiv190708395H} models. We show the MFS unbinned and binned $V_{\mathrm{MFS}}$ values for each source as well as the predicted $V_{\mathrm{MFS,max}}$ after a year of observations and the predicted time scale of variability using a central frequency of 1284\,MHz.}}
    \label{tab: RISS predications}
\end{table*}

\begin{figure*}
\begin{center}
\includegraphics[width=\textwidth]{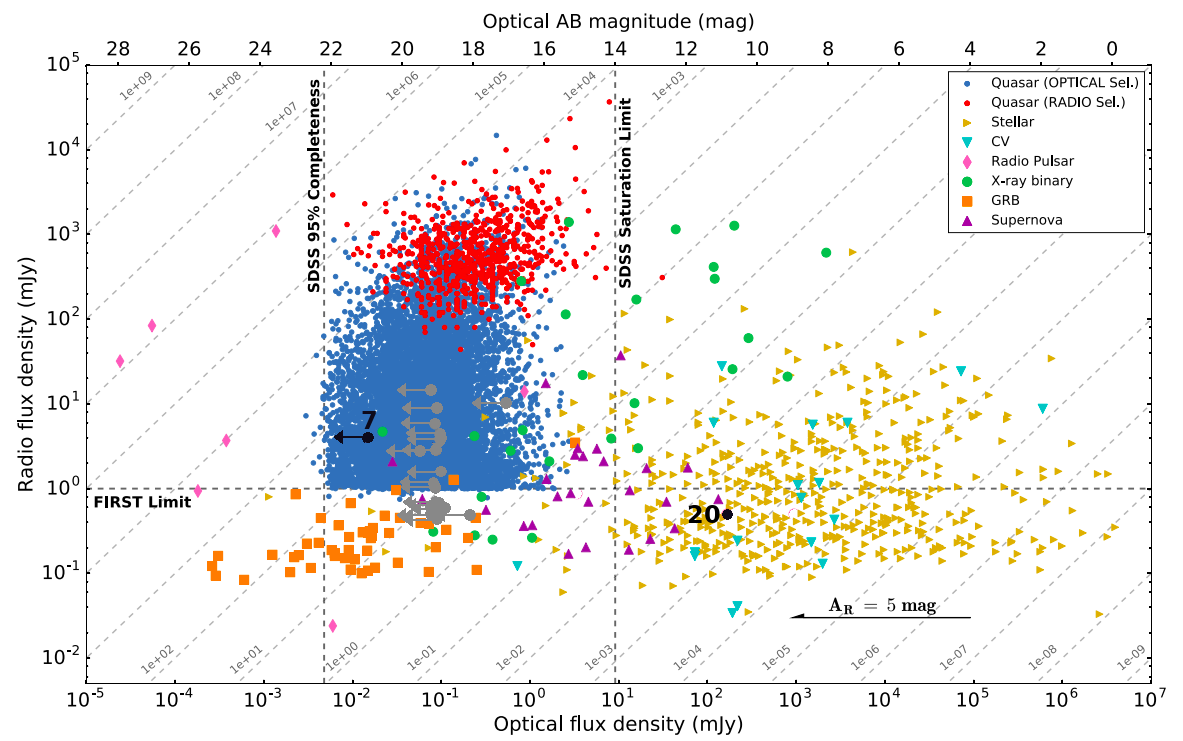}
\caption{Plot from \citet{2018MNRAS.479.2481S} showing the radio and optical flux densities of various stellar sources.
The long-term variable sources have been included as grey markers where the radio flux densities are the median flux densities and the optical flux densities are the extinction-corrected MeerLICHT $r$-band upper limits,
except for 
Source \fbnumba\,which is known stellar flaring system \fb\,(black marker).
Source \gnumba\,(\gname) is also marked in black.
{The extinction values were determined using the NASA/IPAC Extragalactic Database Coordinate Transformation and Galactic Extinction Calculator.}}
\label{fig: Stewart plot}
\end{center}
\end{figure*}

{Most of the new long-term variable sources that we have identified with MeerKAT have measured properties that are consistent with scintillating AGNs.} In Figure\,\ref{fig: Stewart plot}, we have placed the sources onto a plot by \citet{2018MNRAS.479.2481S} that shows the radio and optical flux densities of a range of different source types. This plot shows that the 21 new long-term variable sources are consistent with quasars. 
We note that there are many quasars below the source detection threshold of the FIRST survey, indicated with a dashed line at 1 mJy, down to the lowest-flux variables in our sample. X-ray binaries (XRBs) and $\gamma$-ray bursts (GRBs) also occupy a similar phase-space to quasars in this plot, meaning that our sources are broadly consistent with the radio to optical brightness ratios of XRBs and GRBs.
However, typical radio light curves at MeerKAT observing frequencies from GRBs and XRBs have very different morphologies and timescales than the variable sources in our sample \citep[e.g.][]{2012ApJ...746..156C,2019MNRAS.489.4836F}. This field is also regularly monitored in the X-ray due to \gx, which would have revealed the presence of more XRBs.
The optical to radio flux density ratios of the sources are also broadly consistent with pulsars in Figure\,\ref{fig: Stewart plot}.
While some individual pulsars have spectral indices as low as -3 and as high as 0, most pulsars have spectral indices between approximately -2.2 to -1.0 with a weighted mean of $-1.60\pm0.03$ \citep{2018MNRAS.473.4436J}. 
Only one new source has a spectral index consistent with the steep spectral indices expected for pulsars: \sname\,(source \snumba,  $\oline{\alpha}=-1.46\pm0.05$). If we exclude source \snumba, the spectral indices of the 21 new variable sources range from $\sim-0.6$ to $\sim1.3$, which does not match with the expected spectral index range for pulsars.
This field has been searched by multiple pulsar surveys including the Parkes Southern Pulsar Survey \citep{1996MNRAS.279.1235M}, the Parkes Multibeam Pulsar Survey \citep[PMPS;][]{2001MNRAS.328...17M}, and, most recently, the High-Time Resolution South survey \citep[HTRU-S;][]{2010MNRAS.409..619K} where part of the field (down to a Galactic latitude of -3.5\,degrees) was covered by the low-latitude survey and part was covered by the medium latitude survey. All of our long-term variable sources fall into the region of the sky covered by the medium latitude survey. Initially we conservatively use the worst sensitivity, which is for short-period pulsars (period $\lesssim$1\,ms), is 0.6\,mJy. The HTRU-S survey searched with a central frequency of 1352\,MHz, corresponding to the 1337\,MHz MeerKAT subband. Using the mean flux densities of each source in the 1337\,MHz subband (see Table\,\ref{tab: VAR supp source summary}) we find that 14 of the 21 new long-term variable sources have flux densities greater than the 0.6\,mJy sensitivity limit of HTRU-S. The best sensitivity of HTRU-S mid-latitude is $\sim$0.3\,mJy for longer period pulsars (period $\gtrsim$1\,ms), and all of the new long-term variable sources have mean flux densities $\gtrsim$0.3\,mJy in the 1337\,MHz band. If all of the new long-term variable sources were pulsars, we would expect that more than half of them would have been discovered previously by the HTRU-S mid-latitude survey. So the combination of spectral index and flux density indicate that it is unlikely that many, if any, of these sources are pulsars. {However, if the sources are, for example, intermittent emitters or eclipsing binaries, this may be why they were missed in previous pulsar surveys \citep[e.g.][]{2019ApJ...884...96K}. Further observations, particularly polarisation observations or targeted pulsar searches, would determine whether the properties of these sources are consistent with pulsars.}

{Using the RISS predicted $V_{\mathrm{MFS}}$ and time scales, the radio-optical flux density ratios, the spectral indices and the light curve shapes, we find that the new long-term variable sources are consistent with refractive scintillation of AGN. \rname\,(source \rnumba) has a marginally higher than predicted $V_{\mathrm{MFS}}$ value, but the variability time scale is consistent with the predicted time scale. Also the RISS model we used predicts the $V_{\mathrm{MFS}}$ after one year of observations, while we have observed our sources for two years. The radio-optical flux density ratio of the source is consistent with AGN, and the light curve shape and spectral index are not consistent with pulsars, gamma-ray bursts, or relativistic binaries. \sname\,(source \snumba) has a $V_{\mathrm{MFS}}$ and variability time scale consistent with RISS, and its radio-optical flux density ratio is consistent with being an AGN. Similar to source \rnumba\,the light curve shape is not consistent with  gamma-ray bursts or relativistic binaries. It does have a steep spectral index that is consistent with what is expected for pulsars; however, AGN can also have steep spectral indices and the other properties of the source are consistent with AGN. Further information, such as polarisation measurements, will be useful in confirming the AGN classification of these sources.
The polarisation properties away from the phase centre for MeerKAT are still being characterised and data reduction techniques are being developed. As such, we are presently not able to determine the polarisation properties of these sources with these data. This will be pursued in the future. Work is ongoing to characterise there sources at higher frequencies with ATCA and polarisation analysis will be part of that work.}

As many of these sources do not have multi-wavelength counterparts, further investigation in the radio is the most promising for determining the nature of these sources. We have successfully proposed for ATCA time to observe the variable sources at 5\,GHz and 9\,GHz to obtain polarisation and further spectral information, these observations took place in mid-2021. Circular polarisation of these sources would point towards stellar or pulsar sources, as opposed to AGN. Deep infrared observations and deeper optical images may also reveal counterparts to some sources. MeerLICHT continues to co-observe the field during the night, and will soon have individual epoch information available to further investigate the positions of these sources in the optical as well as the potential to stack more images for greater sensitivity.  MeerTRAP also commensally observes the field while ThunderKAT is observing, which means that we are searching for any short timescale bursts or pulsations from these sources, indicating pulsars or rotating radio transients. ThunderKAT will continue to observe the field weekly until September 2023, providing further radio information about these sources. 

\section{Conclusions}
\label{sec: VAR conclusions}

We present the radio light curves and spectral indices of 21 new long-term variable sources discovered and monitored commensally in two years of MeerKAT observations of the low mass X-ray binary \gx. This is an unprecedented data set in terms of cadence, continuity, and length. Three of these new sources are coincident with counterparts at other frequencies, and one of those is coincident with a MeerLICHT optical source. We find that refractive scintillation of AGNs could explain the long-term variability that we see in these sources. The \gx\,field will continue to be monitored weekly with MeerKAT until September 2023, MeerLICHT will continue to monitor the field during night time observations, and we will use ATCA observations to further investigate the sources.

\section*{Acknowledgements}

LND and BWS acknowledge support from the European Research Council (ERC) under the European Union's Horizon 2020 research and innovation programme (grant agreement No 694745).
ET acknowledges financial support from the UnivEarthS Labex program of Sorbonne Paris Cit\'{e} (ANR-10-LABX-0023 and ANR-11-IDEX-0005-02).
PAW acknowledges support from the National Research Foundation (NRF) and the University of Cape Town (UCT).
A.H. acknowledges support by the I-Core Program of the Planning and Budgeting Committee and the Israel Science Foundation, and support by ISF grant 647/18.This research was supported by a Grant from the GIF, the German-Israeli Foundation for Scientific Research and Development. A.H. Acknowledges support from GIF. This research was supported by Grant No. 2018154 from the United States-Israel Binational Science Foundation (BSF).
This work is based on the research supported in part by the National Research Foundation of South Africa (Grant Numbers 93405 and 119446).
We acknowledge use of the Inter-University Institute for Data Intensive Astronomy (IDIA) data intensive research cloud for data processing. IDIA is a South African university partnership involving the University of Cape Town, the University of Pretoria and the University of the Western Cape.
The MeerKAT telescope is operated by the South African Radio Astronomy Observatory (SARAO), which is a facility of the National Research Foundation, an agency of the Department of Science and Innovation.
We would like to thank the operators, SARAO staff and ThunderKAT Large Survey Project team.
This research made use of Astropy,\footnote{\href{http://www.astropy.org}{http://www.astropy.org}} a community-developed core Python package for Astronomy \citep{2013A&A...558A..33A,2018AJ....156..123A}.
This research made use of APLpy, an open-source plotting package for Python \citep{2012ascl.soft08017R}.
LND would like to thank Tiaan Bezuidenhout, Manisha Caleb, Fabian Jankowski, Mat Malenta, Vincent Morello, Kaustubh Rajwade, and Mayuresh Surnis for useful and interesting discussions.

\section*{Data availability}

The data underlying this article are available in Zenodo at \href{http://doi.org/10.5281/zenodo.5069119}{http://doi.org/10.5281/zenodo.5069119}. 
The code underlying this article is available on Zenodo: \href{https://doi.org/10.5281/zenodo.4456303}{https://doi.org/10.5281/zenodo.4456303}
and
\href{https://doi.org/10.5281/zenodo.4921715}{https://doi.org/10.5281/zenodo.4921715}.




\bibliographystyle{mnras}
\bibliography{LongTermVariables} 



\appendix

\section{Flux density systematics}
\label{app: systematics}

\begin{figure}
\begin{center}
	\includegraphics[width=\columnwidth]{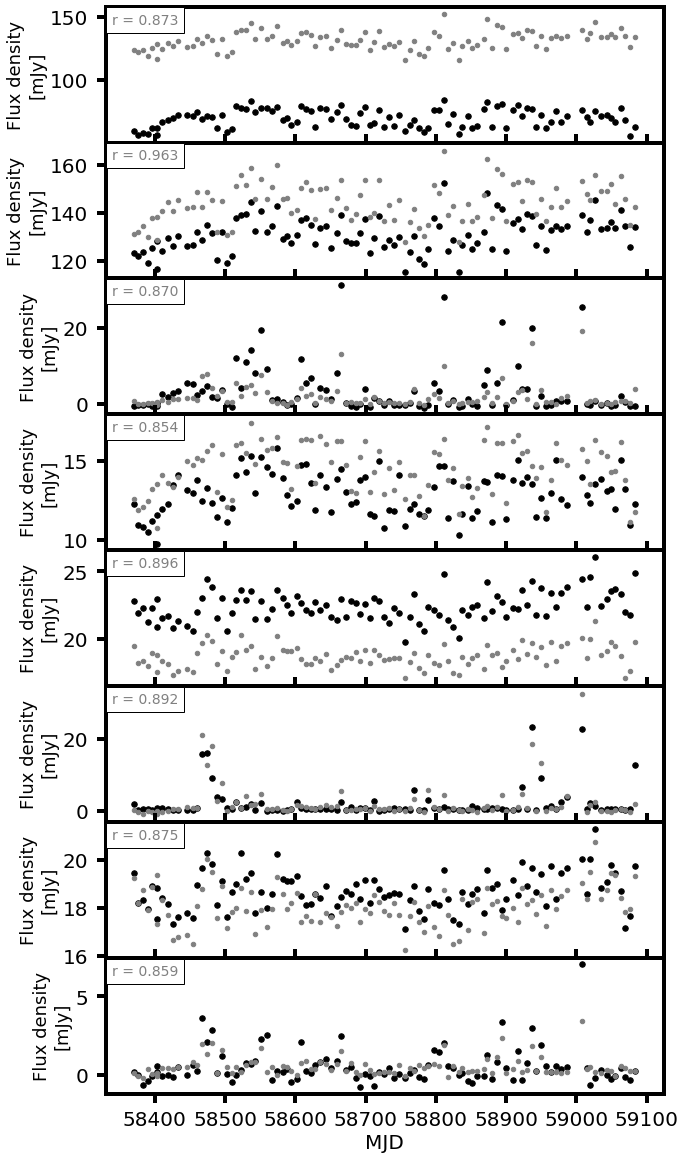}
    \caption[Examples of highly correlated light curves.]{Examples of pairs of highly correlated light curves. Each row shows light curves from two different sources, one in grey and one in black. The Pearson's \pcc\,value calculated by comparing each pair of light curves is shown in the upper left corner of each panel.}
    \label{fig: correlated sources examples}
\end{center}
\end{figure}

Upon visual inspection of the light curves of sources in the \gx\,field, we noticed that some show correlated variability. Some examples of source light curves are shown in Figure\,\ref{fig: correlated sources examples}. This is concerning for variability searches as these sources appear variable, but are unlikely to be as they are apparently varying in the same way. This could also indicate underlying issues that mean that every source in this field may be affected by this systematic variability. To investigate the correlated sources further and determine the cause of the correlation we started with the Pearson's $r$ correlation coefficient \citep[\pcc;][]{1896RSPTA.187..253P}. Pearsons's \pcc\,measures the linear correlation between two sets of discrete points, where an \pcc\,of $-1$ is a $100\%$ negative correlation, an \pcc\,of $+1$ is a $100\%$ positive correlation, and an \pcc\,of $0$ is no correlation at all. The \pcc\,values between the sources shown in Figure\,\ref{fig: correlated sources examples} are all above 0.85. We can see similar peaks and troughs, as well as overall, long-term trends. When we compare the light curves of all sources in the field with each other (only once, if we calculate the \pcc\,comparing source A to source B we did not calculate the \pcc\,comparing source B to source A), we find that many sources correlate strongly with each other.

\subsection{Resolved sources and artefacts}
\label{sec: SYS removing resolved}

Our first attempt to mitigate the correlation problem was to vet all sources to check for artefacts and resolved sources. The \gx\,field does not have any large extended structures, the largest resolved structures are lobes of radio galaxies and a planetary nebula. The {\tt force\_beam} and {\tt beamwidths\_limit} parameters mean that obvious extended sources are not included by \trap. There are many double-lobed sources and small resolved sources and sources with structure that can appear point-like. It is important to remember that these are images with integration times of 10-15\,minutes, which means that the synthesised beam is not round.  Sometimes two sources that are close together on the sky can appear to be a single point source depending on the orientation of the beam. This means that
flux is sometimes collected from mostly one source, or sometimes both sources. There are also clean artefacts around bright sources that are detected as point-like sources in some epochs. These artefacts vary from epoch to epoch and often appear and disappear completely, which means that they can look like very interesting variable sources.

Differentiating between resolved and unresolved sources in radio images is challenging. As we are investigating a single field, we chose a manual, visual method to remove the problem sources. To do this, we created a DS9 \citep{SAODS9} region file including every source detected by \trap\,in every subband and the MFS images. We then looked at these sources in the deep MeerKAT stacked image of this field and determined by-eye which sources are resolved and which sources are artefacts. We then checked all the sources again using a single-epoch 1658\,MHz image as sources are more likely to be resolved at the highest frequency due to the smaller synthesised beam. We then removed these sources and compared all the sources in the field to each other using the Pearson's \pcc\,coefficient. This method removes all but one of the light curves from Figure\,\ref{fig: correlated sources examples}, meaning that all but one of these sources is either a resolved source or an artefact, and reduces the amount of strong correlations within the \gx\,field.

\subsection{Direction dependence of the systematics}
\label{sec: SYS direction dependence}

The next possible cause of the correlation that we tested was direction dependence, in particular distance from the phase centre (centred on \gx).
To do this, we select all the sources within an annulus. We then choose a source within that annulus, and calculate the \pcc\,correlation coefficient between that source and all of the other sources within the annulus. We then remove that source, so that we do not compare to it again. We proceed to do this for every source. We perform the same calculations for annuli outwards from the phase centre. We choose the radii of each annulus such that the area of every annulus (and the corresponding circle located at the phase centre) is the same. We find that there is no clear direction-dependence, no annulus where the correlation is particularly strong.

\subsection{Flux dependence of the systematics}
\label{sec: SYS flux dependence}

Finally, we wanted to determine whether the systematics are flux density dependent. In other words: are brighter sources more affected by the systematics? To do this, we compared all the sources using the \pcc\,correlation coefficient again and plotted the flux densities and correlation coefficient value range. We find that there may be some flux density dependence where bright sources are more affected by the systematics than fainter sources. This could be because the effect is multiplicative, or it could be because the uncertainties on bright sources are underestimated by \trap\,(Rowlinson, A., private communications).

\subsection{Modelling the light curve systematics}
\label{sec: SYS median correction}

Removing resolved sources and artefacts, see Section\,\ref{sec: SYS removing resolved}, reduces the correlation between sources, but some residual correlation remains.  As such we developed a method to model the light curve systematics.
To do this, we first choose a reference epoch. We use the last epoch as the reference epoch, because of the way \trap\, tracks sources. We force \trap\, to track all detected sources in every epoch, therefore all sources will have a measured value from \trap\, in the last epoch. We then take every light curve (where the source is detected at least once with a S/N$>3$) and divide all of the epochs by the reference epoch value:
\begin{eqnarray}
    F_{i, j, \mathrm{scaled}} = \frac{F_{i,j}}{F_{i, \mathrm{ref\,epoch}}}
\end{eqnarray}
where $F_{i,j}$ is the flux density of source $i$ in epoch $j$ and $F_{i, \mathrm{ref\,epoch}}$ is the flux density of the same source in the reference epoch.
Next, we take the mean ($\oline{F}$), standard deviation ($\sigma_{F}$), median ($\widetilde{F}$), and MAD ($\mathrm{MAD}_{F}$) of this distribution for every epoch, where the mean is given by:
\begin{eqnarray}
    \oline{F} =  \frac{1}{n} \sum^{n}_{i=0}F_{i}
\end{eqnarray}
where $n$ is the number of epochs. The standard deviation is given by:
\begin{eqnarray}
    \sigma_{F} = \sqrt{\frac{1}{n\cdot \left(n-1\right)}\sum^{n}_{i=0}\left( F_{i} - \oline{F} \right)^{2}} .
\end{eqnarray}
The median ($\widetilde{F}$) is the value where half of the population is less than the value of the median
and the MAD is defined by:
\begin{eqnarray}
    \mathrm{MAD} = \mathrm{median}\left(\left| F_{i} - \widetilde{F} \right| \right) .
    \label{eq: MAD definition}
\end{eqnarray}
Here, we use the MAD value multiplied by $1.4826/\sqrt{n}$, such that $\mathrm{MAD}_{F} = \frac{\mathrm{MAD}\cdot 1.4826}{\sqrt{n}}$ where $1.4826$ is the scale factor to use the MAD as an estimation of the standard deviation for normally distributed data, and the $1/\sqrt{n}$ factor makes the MAD comparable to the population standard deviation. We can now use either the mean and standard deviation per epoch or the median and MAD per epoch as the model for the systematics. 

\begin{figure}
\begin{center}
	\includegraphics[width=\columnwidth]{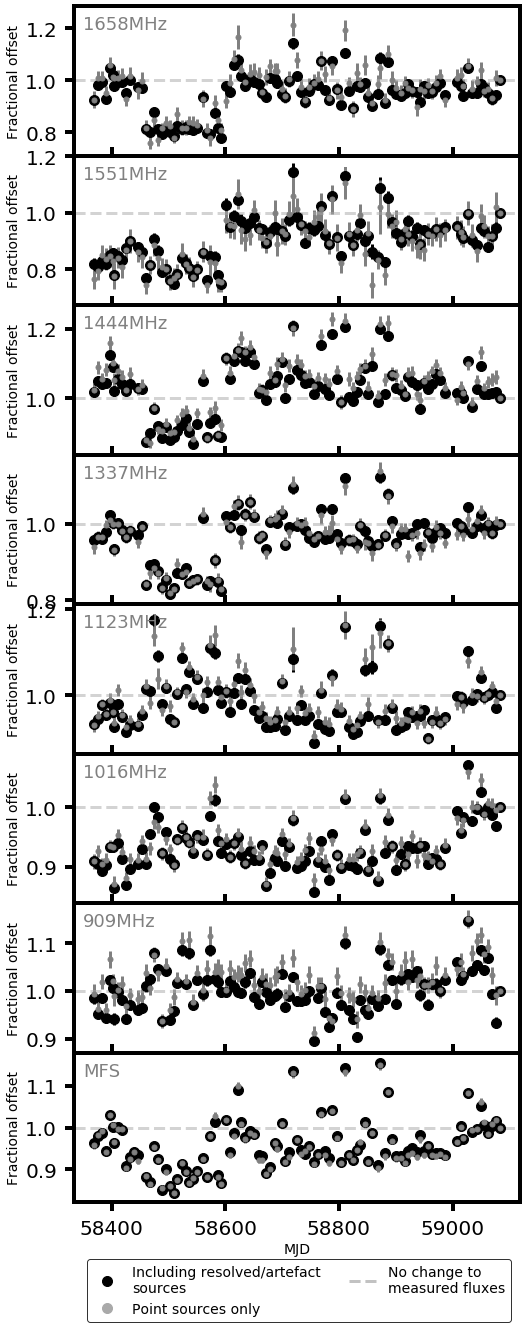}
    \caption[Scale models for all frequency subbands.]{Median and MAD scale models for all frequency subbands and MFS. All points have errorbars, some are too small to see. The grey-dashed line is at 1.0.}
    \label{fig: all model demo nocut}
\end{center}
\end{figure}

We use the median and MAD model to model and correct for the systematics as the median is robust to outliers compared to the mean. The median model for each frequency band and the MFS images is shown in Figure\,\ref{fig: all model demo nocut} (left column). This plot also shows the difference between the models before and after removing the resolved sources and artefacts. This shows that removing these sources does not significantly change the shape of the models, indicating that the main contribution to the systematics is not these sources. To correct for the systematics, we divide each light curve by the model and combine the uncertainties.

\subsection{Systematic correction results}
\label{sec: SYS systematics results}

\begin{figure*}
\begin{center}
	\includegraphics[width=\textwidth]{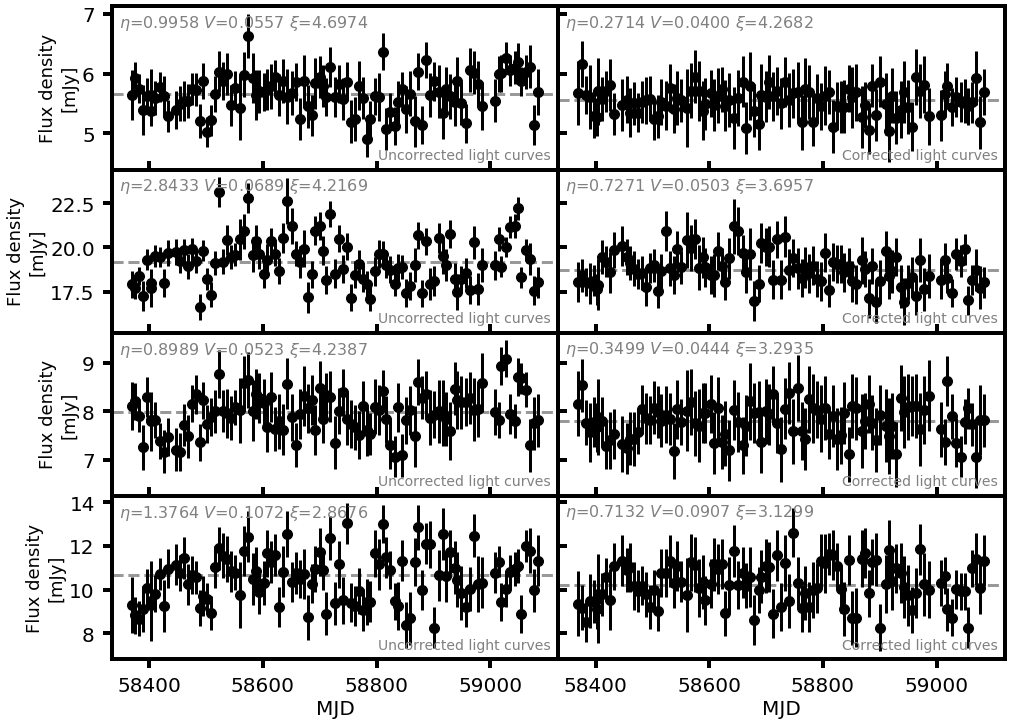}
    \caption[Examples of light curves before and after correction.]{Examples of 909\,MHz subband light curves before and after correction. Each row is the same source, where the left column is the light curve before the correction is applied, and the right column is the light curve after the correction has been applied. These sources are all point sources with a S/N\,$>2$.}
    \label{fig: uncorr vs corr example light curves}
\end{center}
\end{figure*}

\begin{figure*}
\begin{center}
	\includegraphics[width=0.75\textwidth]{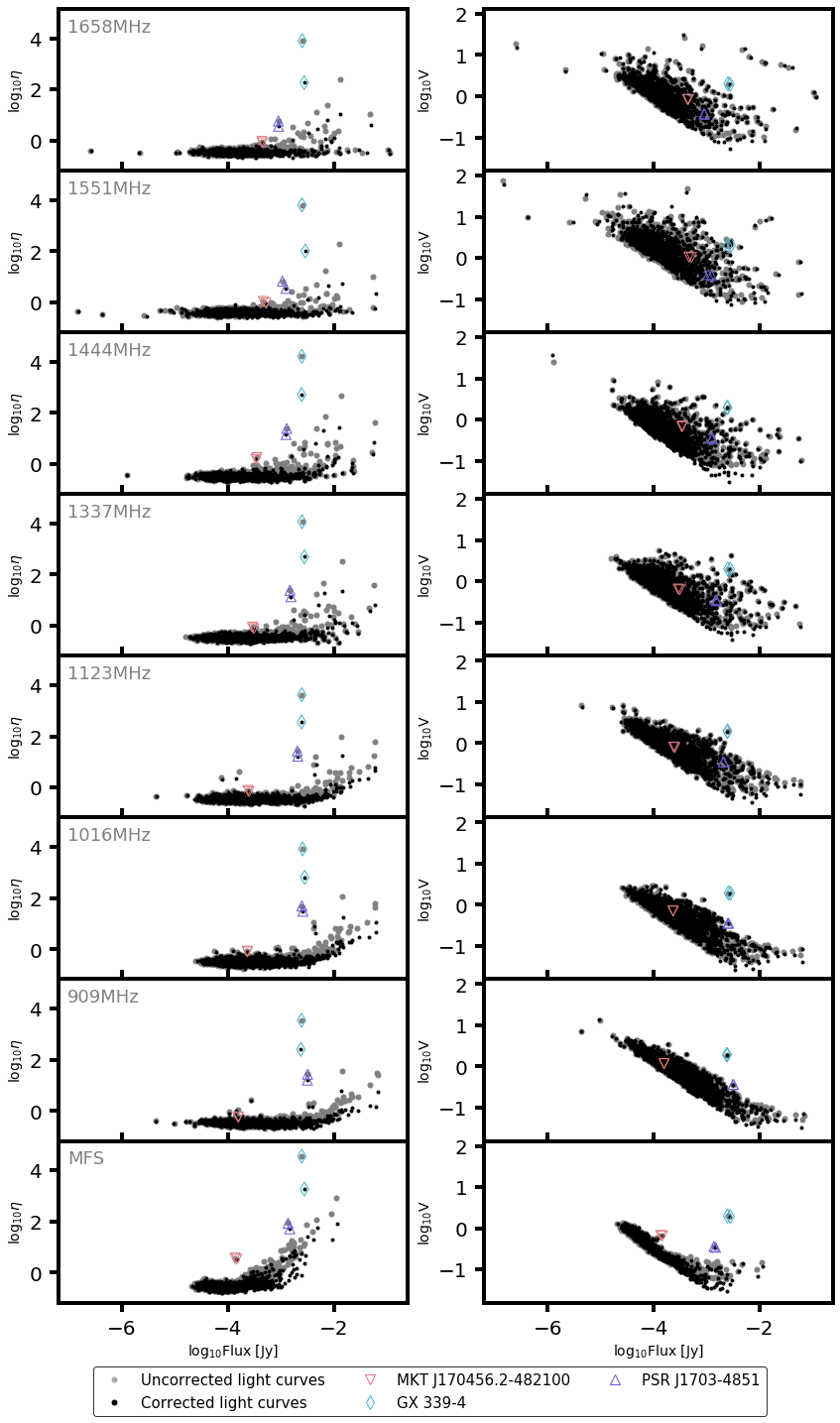}
    \caption[Variability parameters before and after applying the corrections.]{Variability parameters before and after applying the corrections. Resolved sources and artefacts have already been removed, as have sources with a S/N\,$<2$.}
    \label{fig: varparams uncorr vs corr}
\end{center}
\end{figure*}

Examples of light curves before and after correction are shown in Figure\,\ref{fig: uncorr vs corr example light curves}. We can see that these sources are consistent with constant sources, and that their variability parameters decrease after applying the corrections, particularly the \varp\,parameter. In Figure\,\ref{fig: varparams uncorr vs corr} we can see the variability parameters before and after correction. These values are for light curves with S/N\,$>2$ in at least one epoch. The code for the investigation and correction of the systematics can be found on GitHub: \href{https://doi.org/10.5281/zenodo.4456303}{https://doi.org/10.5281/zenodo.4456303}. This repository also includes the code for accessing the light curves from \trap.

\section{Supplementary material}
\label{sec: VAR supp source info}

\begin{figure*}
\includegraphics[width=\textwidth]{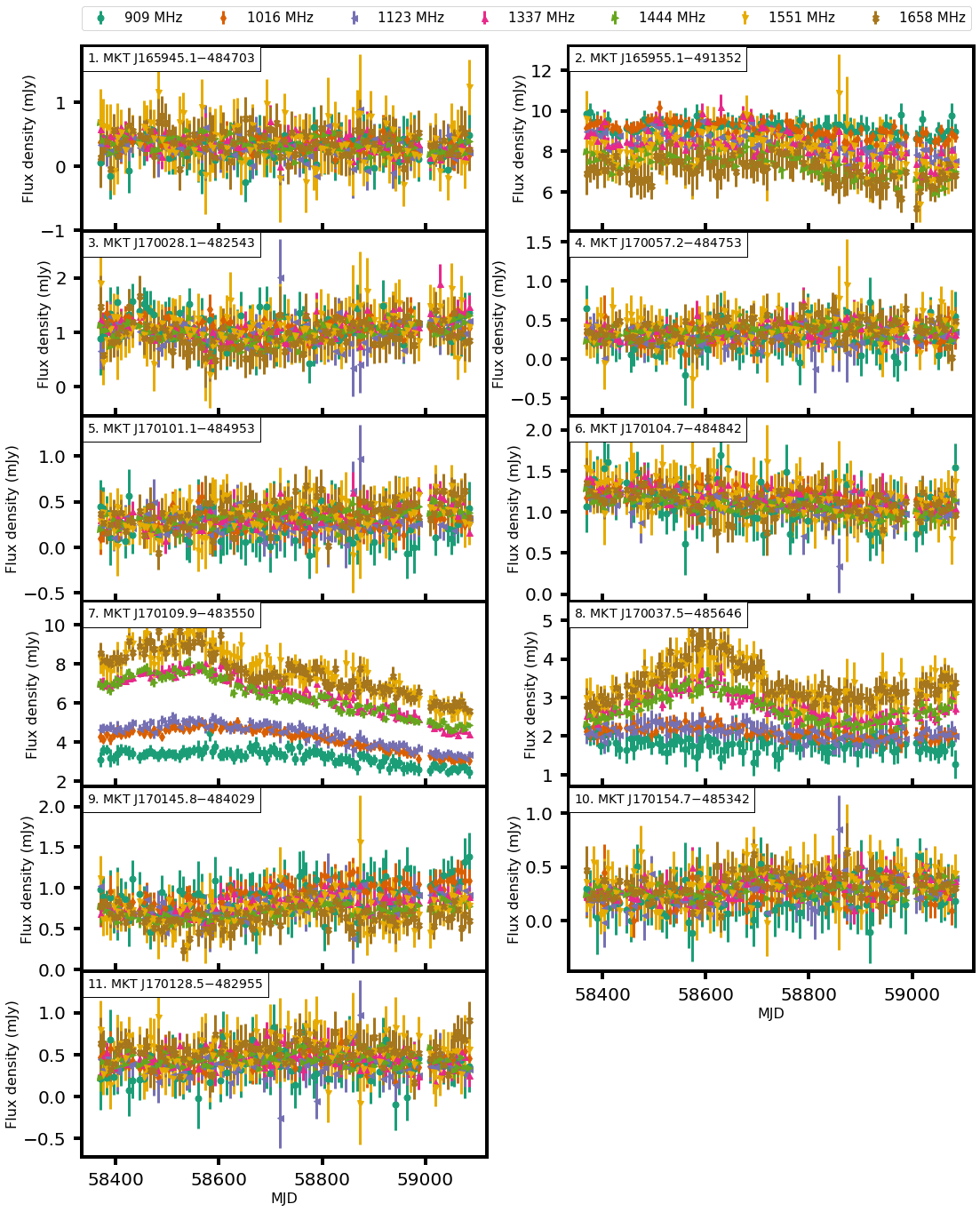}
\caption{Subband light curves for sources: 
(\anumba) \aname,
(\bnumba) \bname,
(\cnumba) \cname,
(\dnumba) \dname,
(\enumba) \ename,
(\fnumba) \fname,
(\gnumba) \gname,
(\hnumba) \hname,
(\inumba) \iname,
(\jnumba) \jname,
(\knumba) \kname. Note that the 1230\,MHz band has been excluded due to RFI and that the subband flux densities have been primary beam corrected.}
\label{fig: subbands 1-13}
\end{figure*}

\begin{figure*}
\includegraphics[width=\textwidth]{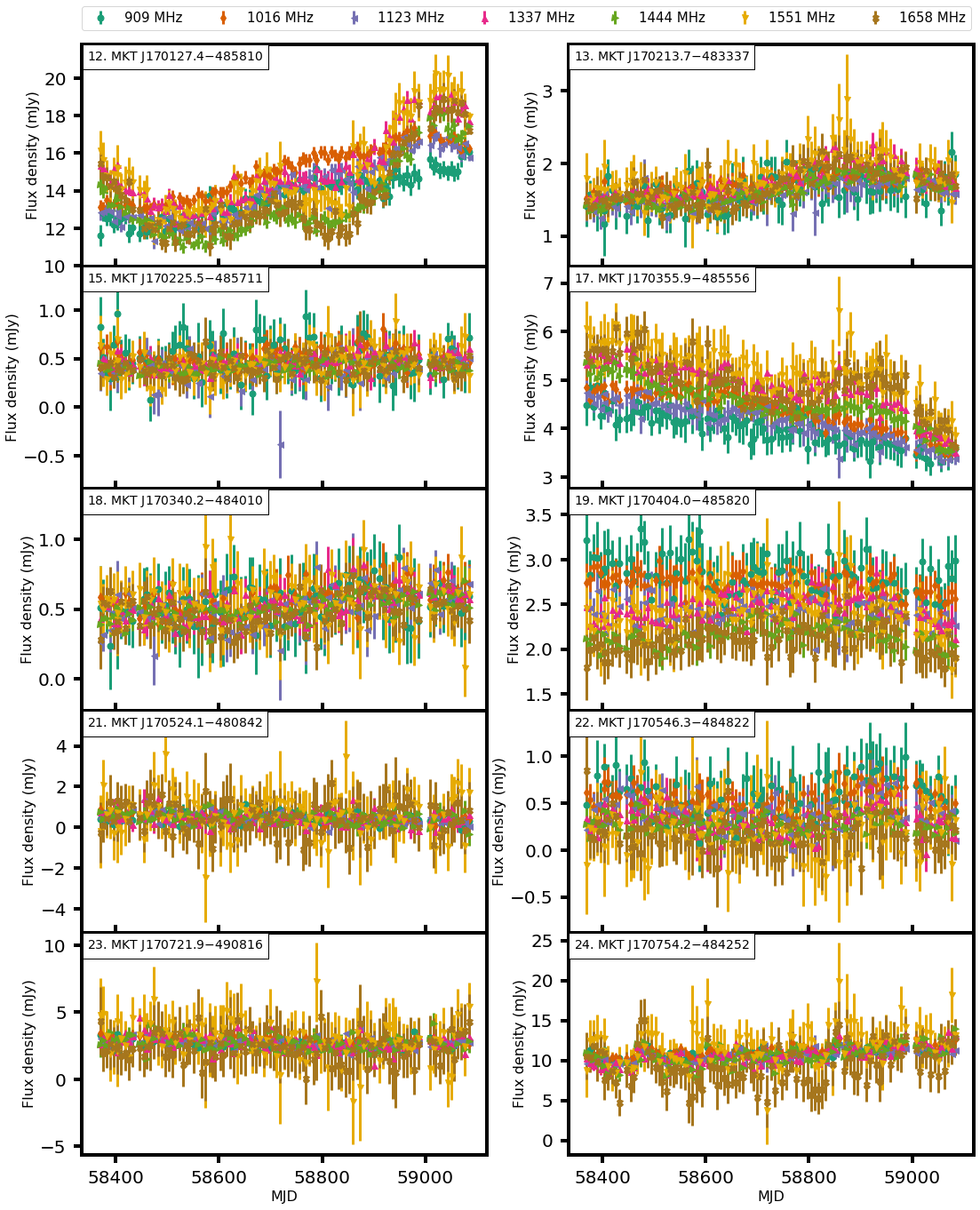}
\caption{Subband light curves for sources:
(\lnumba) \lname,
(\mnumba) \mname,
(\nnumba) \nname,
(\onumba) \oname,
(\pnumba) \pname,
(\qnumba) \qname,
(\rnumba) \rname,
(\snumba) \sname,
(\tnumba) \tname,
(\unumba) \uname. Note that the 1230\,MHz band has been excluded due to RFI and that the subband flux densities have been primary beam corrected.}
\label{fig: subbands 14-28}
\end{figure*}

The mean flux density for each long-term variable source is shown in Table\,\ref{tab: VAR supp source summary}.
The Vizier catalogues used for matching to the long-term variable sources are shown in Table\,\ref{tab: vizier cats}.
Figures\,\ref{fig: subbands 1-13} and \ref{fig: subbands 14-28} show the subband light curves for the 21 new long-term variables sources.

\begin{landscape}
\begin{table}
\centering
\begin{tabular}{llrr|rrrrrrr}
\hline
    &     &    &     &  \multicolumn{7}{c}{Mean flux density (mJy) in subband (MHz)} \\
    \hline
    & Name & RA (deg) & Dec (deg) & 1658 & 1551 & 1444 & 1337 & 1123 & 1016 & 909 \\
    \hline
    \hline
   \anumba & \aname & 254.93796 & $-$48.78424 & $0.37\pm0.03$ & $0.35\pm0.04$ & $0.34\pm0.02$ & $0.37\pm0.02$ & $0.30\pm0.02$ & $0.35\pm0.02$ & $0.29\pm0.03$ \\
    \bnumba & \bname & 254.97960 & $-$49.23137 & $7.13\pm0.08$ & $7.9\pm0.1$ & $7.33\pm0.04$ & $8.27\pm0.04$ & $8.34\pm0.04$ & $9.09\pm0.03$ & $9.13\pm0.05$ \\
    \cnumba & \cname & 255.11701 & $-$48.42868 & $0.93\pm0.04$ & $1.03\pm0.04$ & $0.98\pm0.02$ & $1.07\pm0.02$ & $1.00\pm0.02$ & $1.11\pm0.02$ & $1.10\pm0.03$ \\
    \dnumba & \dname & 255.23823 & $-$48.79807 & $0.37\pm0.02$ & $0.37\pm0.02$ & $0.36\pm0.01$ & $0.37\pm0.01$ & $0.29\pm0.02$ & $0.30\pm0.02$ & $0.26\pm0.02$ \\
    \enumba & \ename & 255.25445 & $-$48.83158 & $0.36\pm0.02$ & $0.35\pm0.02$ & $0.33\pm0.01$ & $0.34\pm0.01$ & $0.28\pm0.02$ & $0.28\pm0.02$ & $0.25\pm0.02$ \\
    \fnumba & \fname & 255.26939 & $-$48.81184 & $1.12\pm0.02$ & $1.17\pm0.03$ & $1.06\pm0.01$ & $1.18\pm0.01$ & $1.11\pm0.02$ & $1.19\pm0.02$ & $1.13\pm0.03$ \\
    \gnumba & \gname & 255.29126 & $-$48.59750 & $7.33\pm0.04$ & $7.27\pm0.05$ & $6.20\pm0.02$ & $6.26\pm0.02$ & $4.33\pm0.02$ & $4.14\pm0.02$ & $3.19\pm0.03$ \\
    \hnumba & \hname & 255.15615 & $-$48.94632 & $3.33\pm0.03$ & $3.25\pm0.04$ & $2.70\pm0.02$ & $2.79\pm0.02$ & $2.12\pm0.02$ & $2.11\pm0.02$ & $1.81\pm0.03$ \\
    \inumba & \iname & 255.44078 & $-$48.67495 & $0.66\pm0.02$ & $0.72\pm0.02$ & $0.70\pm0.01$ & $0.79\pm0.01$ & $0.79\pm0.02$ & $0.92\pm0.01$ & $0.90\pm0.02$ \\
    \jnumba & \jname & 255.47793 & $-$48.89498 & $0.32\pm0.01$ & $0.33\pm0.02$ & $0.297\pm0.009$ & $0.31\pm0.01$ & $0.27\pm0.01$ & $0.26\pm0.01$ & $0.23\pm0.02$ \\
    \knumba & \kname & 255.36859 & $-$48.49876 & $0.53\pm0.02$ & $0.49\pm0.03$ & $0.44\pm0.01$ & $0.46\pm0.01$ & $0.38\pm0.02$ & $0.43\pm0.02$ & $0.38\pm0.03$ \\
    \lnumba & \lname & 255.36419 & $-$48.96960 & $13.78\pm0.06$ & $14.87\pm0.08$ & $13.29\pm0.03$ & $15.01\pm0.04$ & $14.00\pm0.04$ & $14.87\pm0.03$ & $13.62\pm0.05$ \\
    \mnumba & \mname & 255.55712 & $-$48.56044 & $1.66\pm0.02$ & $1.78\pm0.03$ & $1.59\pm0.01$ & $1.72\pm0.01$ & $1.59\pm0.02$ & $1.69\pm0.02$ & $1.61\pm0.03$ \\
    \gxnumba & \gx & 255.70547 & $-$48.78973 & $1.37\pm0.02$ & $1.42\pm0.02$ & $1.51\pm0.01$ & $1.69\pm0.01$ & $1.89\pm0.02$ & $2.28\pm0.02$ & $2.28\pm0.03$ \\
    \nnumba & \nname & 255.60623 & $-$48.95290 & $0.42\pm0.02$ & $0.44\pm0.02$ & $0.419\pm0.009$ & $0.47\pm0.01$ & $0.43\pm0.01$ & $0.49\pm0.01$ & $0.48\pm0.02$ \\
    \gxpsrnumba & \gxpsr & 255.97719 & $-$48.86696 & $0.91\pm0.02$ & $1.14\pm0.02$ & $1.19\pm0.01$ & $1.52\pm0.01$ & $2.09\pm0.02$ & $2.56\pm0.02$ & $3.09\pm0.03$ \\
    \onumba & \oname & 255.98298 & $-$48.93234 & $4.86\pm0.03$ & $5.02\pm0.04$ & $4.50\pm0.02$ & $4.81\pm0.02$ & $4.11\pm0.02$ & $4.34\pm0.02$ & $3.98\pm0.03$ \\
    \pnumba & \pname & 255.91730 & $-$48.66967 & $0.48\pm0.02$ & $0.55\pm0.02$ & $0.500\pm0.009$ & $0.53\pm0.01$ & $0.52\pm0.01$ & $0.57\pm0.01$ & $0.54\pm0.02$ \\
    \qnumba & \qname & 256.01674 & $-$48.97230 & $2.09\pm0.02$ & $2.36\pm0.03$ & $2.14\pm0.01$ & $2.43\pm0.02$ & $2.42\pm0.02$ & $2.72\pm0.02$ & $2.80\pm0.03$ \\
    \fbnumba & \fb & 256.23435 & $-$48.35021 & $0.40\pm0.04$ & $0.42\pm0.04$ & $0.36\pm0.02$ & $0.33\pm0.02$ & $0.25\pm0.02$ & $0.25\pm0.02$ & $0.19\pm0.03$ \\
    \rnumba & \rname & 256.35003 & $-$48.14511 & $0.5\pm0.1$ & $0.6\pm0.1$ & $0.47\pm0.05$ & $0.47\pm0.05$ & $0.43\pm0.04$ & $0.48\pm0.03$ & $0.45\pm0.04$ \\
    \snumba & \sname & 256.44297 & $-$48.80631 & $0.20\pm0.03$ & $0.20\pm0.03$ & $0.26\pm0.02$ & $0.32\pm0.02$ & $0.40\pm0.02$ & $0.52\pm0.02$ & $0.55\pm0.03$ \\
    \tnumba & \tname & 256.84123 & $-$49.13783 & $2.3\pm0.2$ & $2.9\pm0.2$ & $2.75\pm0.07$ & $2.82\pm0.07$ & $2.78\pm0.05$ & $2.87\pm0.04$ & $2.71\pm0.05$ \\
    \unumba & \uname & 256.97587 & $-$48.71454 & $9.3\pm0.2$ & $11.4\pm0.2$ & $10.52\pm0.09$ & $10.65\pm0.09$ & $10.49\pm0.06$ & $11.17\pm0.05$ & $11.03\pm0.06$ \\
    \hline
\end{tabular}
\caption[]{Summary of the positions and mean flux densities of the long-term variable sources in the \gx\,field. The RA and Dec are in degrees and have been corrected for the absolute astrometry (see Section\,\ref{sec: VAR absolute astrometry}), both the RA and Dec have uncertainties of 0\farcs4.}
\label{tab: VAR supp source summary}
\end{table}
\end{landscape}

\begin{figure*}
\begin{center}
\includegraphics[width=\textwidth]{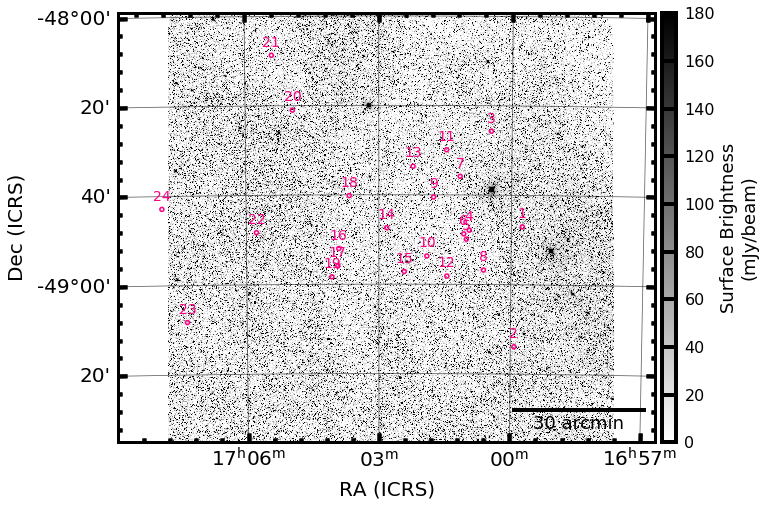}
\caption[Deep MeerLICHT $q$-band image including the long-term variable sources.]{Deep MeerLICHT $q$-band image including the long-term variable sources.}
\label{fig: VAR SUPP MLT image}
\end{center}
\end{figure*}

\begin{figure*}
\begin{center}
\includegraphics[width=\textwidth]{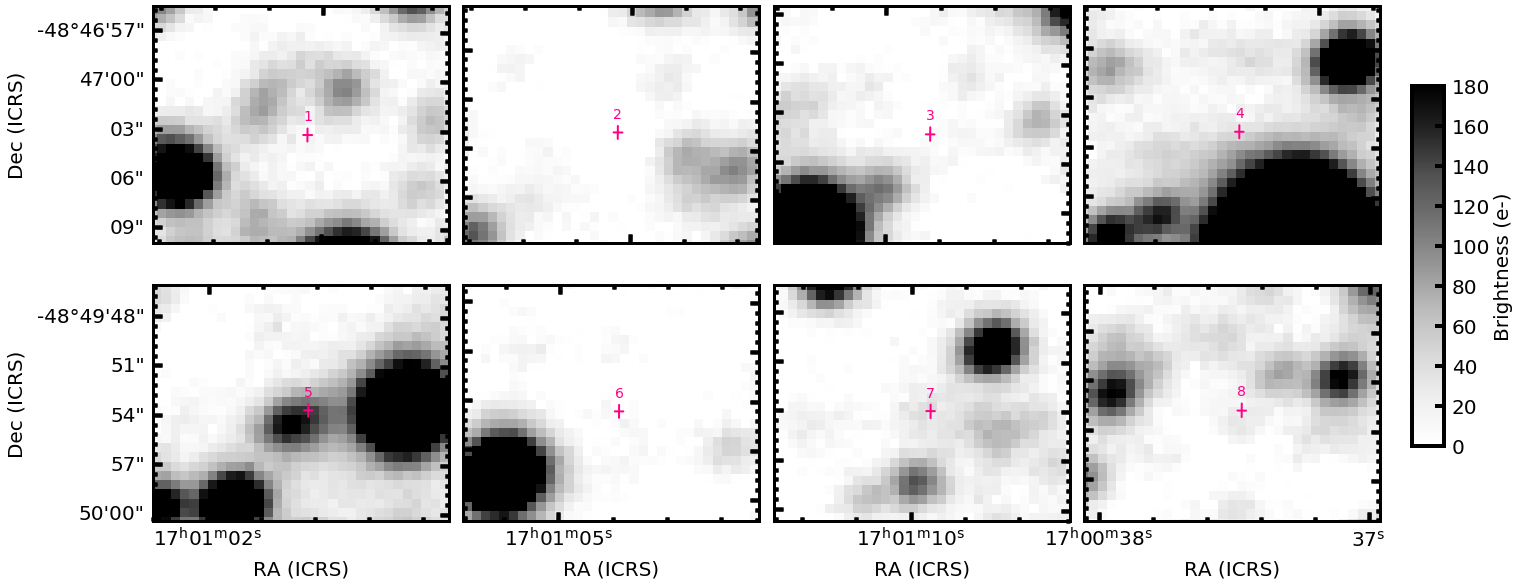}
\caption{Deep MeerLICHT $q$-band postage stamps showing the locations of sources 1 to 8. {The cross-hairs are 0\farcs4 long to indicate the astrometric uncertainties on the MeerKAT positions.} These sources are:
(\anumba) \aname; 
(\bnumba) \bname; 
(\cnumba) \cname; 
(\dnumba) \dname; 
(\enumba) \ename; 
(\fnumba) \fname;
(\gnumba) \gname;
and (\hnumba) \hname.}
\label{fig: VAR SUPP MLT stamps 1-8}
\end{center}
\end{figure*}

\begin{figure*}
\begin{center}
\includegraphics[width=\textwidth]{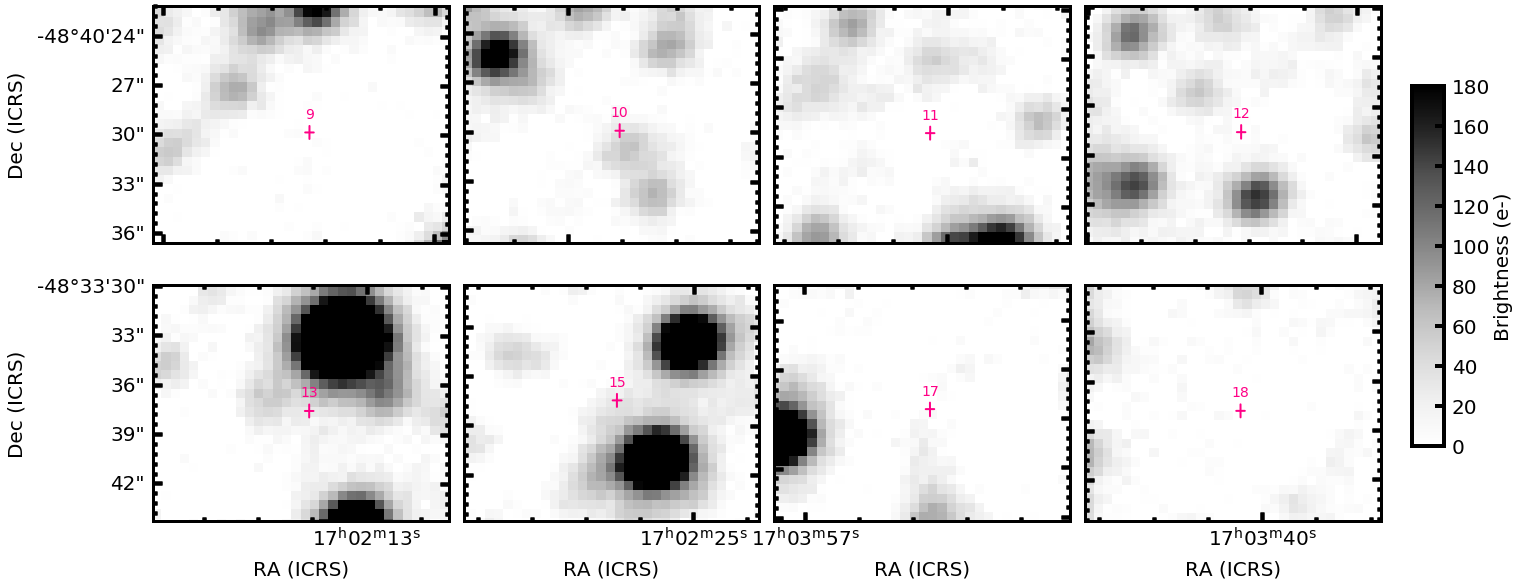}
\caption{Deep MeerLICHT $q$-band postage stamps showing the locations of sources 9 to 16. {The cross-hairs are 0\farcs4 long to indicate the astrometric uncertainties on the MeerKAT positions.} These sources are: 
(\inumba) \iname; 
(\jnumba) \jname; 
(\knumba) \kname; 
(\lnumba) \lname; 
(\mnumba) \mname; 
(\nnumba) \nname; 
(\onumba) \oname; and 
(\pnumba) \pname.
}
\label{fig: VAR SUPP MLT stamps 9-16}
\end{center}
\end{figure*}

\begin{figure*}
\begin{center}
\includegraphics[width=\textwidth]{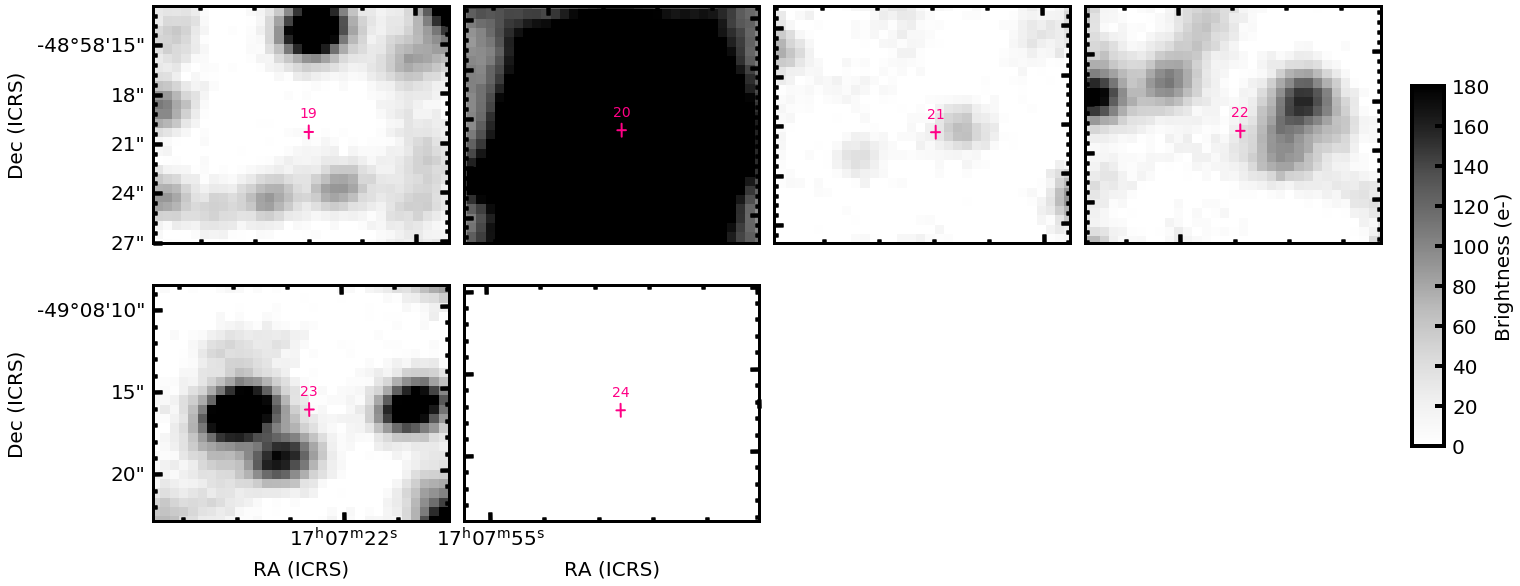}
\caption{Deep MeerLICHT $q$-band postage stamps showing the locations of sources 17 to 24. {The cross-hairs are 0\farcs4 long to indicate the astrometric uncertainties on the MeerKAT positions.} These sources are: 
(\qnumba) \qname; 
(\fbnumba) \fb; 
(\rnumba) \rname; 
(\snumba) \sname; 
(\tnumba) \tname; and 
(\unumba) \uname.
As source \fbnumba\,is \fb\,the star is overexposed, leading to the large black blob we can see here.}
\label{fig: VAR SUPP MLT stamps 17-24}
\end{center}
\end{figure*}

The MeerLICHT $5\sigma$ upper limits for the location of each source are shown in Table\,\ref{tab: VAR supp MLT lims}. One of the MeerLICHT products is a FITS image map of the limiting magnitude across the whole MeerLICHT field. These limits were measured at the position of each source in the limit maps per band. The MeerLICHT $q$-band deep image with the locations of the variable sources (numbers correspond to those in Table\,\ref{tab: VAR supp source spec summary}) are shown in Figure\,\ref{fig: VAR SUPP MLT image}. Postage stamps of the positions of each long-term variable source in the MeerLICHT deep $q$-band are shown in Figures\,\ref{fig: VAR SUPP MLT stamps 1-8} to \ref{fig: VAR SUPP MLT stamps 17-24}.

\begin{table*}
\centering
\begin{tabular}{llrrrrrr}
     & Name & $u$ & $g$ & $r$ & $i$ & $z$ & $q$ \\
    \hline
    \hline
   \anumba & \aname & 21.07  /  17.19 & 21.58  /  18.56 & 21.24  /  19.15 & 21.10  /  19.57 & 19.72 / 18.56 & 22.37 / 19.82 \\
    \bnumba & \bname & 21.19  /  17.31 & 21.74  /  18.72 & 21.44  /  19.35 & 21.46  /  19.93 & 19.87 / 18.71 & 22.76 / 20.21 \\
    \cnumba & \cname & 21.23  /  17.35 & 21.60  /  18.58 & 21.24  /  19.15 & 21.44  /  19.91 & 19.74 / 18.58 & 22.62 / 20.07 \\
    \dnumba & \dname & 20.82  /  16.94 & 21.43  /  18.41 & 21.07  /  18.98 & 20.84  /  19.31 & 19.53 / 18.37 & 22.02 / 19.47 \\
    \enumba & \ename & 21.16  /  17.28 & 21.17  /  18.15 & 20.51  /  18.42 & 19.94  /  18.41 & 19.03 / 17.87 & 21.29 / 18.74 \\
    \fnumba & \fname & 21.09  /  17.21 & 21.64  /  18.62 & 21.36  /  19.27 & 21.29  /  19.76 & 19.79 / 18.63 & 22.58 / 20.03 \\
    \gnumba & \gname & 20.82  /  16.94 & 21.51  /  18.49 & 21.13  /  19.04 & 20.52  /  18.99 & 19.45 / 18.29 & 22.06 / 19.51 \\
    \hnumba & \hname & 21.37  /  17.49 & 21.84  /  18.82 & 21.38  /  19.29 & 21.03  /  19.50 & 19.93 / 18.77 & 22.35 / 19.80 \\
    \inumba & \iname & 20.94  /  17.06 & 21.62  /  18.60 & 21.34  /  19.25 & 21.40  /  19.87 & 19.74 / 18.58 & 22.68 / 20.13 \\
    \jnumba & \jname & 21.07  /  17.19 & 21.72  /  18.70 & 21.09  /  19.00 & 20.75  /  19.22 & 19.58 / 18.42 & 22.16 / 19.61 \\
    \knumba & \kname & 20.85  /  16.97 & 21.64  /  18.62 & 21.34  /  19.25 & 21.38  /  19.85 & 19.79 / 18.63 & 22.66 / 20.11 \\
    \lnumba & \lname & 21.37  /  17.49 & 21.86  /  18.84 & 21.50  /  19.41 & 21.55  /  20.02 & 19.85 / 18.69 & 22.78 / 20.23 \\
    \mnumba & \mname & 20.89  /  17.01 & 21.47  /  18.45 & 21.17  /  19.08 & 21.27  /  19.74 & 19.74 / 18.58 & 22.24 / 19.69 \\
    \gxnumba & \gx & 20.26  /  16.38 & 20.74  /  17.72 & 20.18  /  18.09 & 19.44  /  17.91 & 18.78 / 17.62 & 20.88 / 18.33 \\
    \nnumba & \nname & 21.39  /  17.51 & 21.62  /  18.60 & 21.13  /  19.04 & 20.75  /  19.22 & 19.72 / 18.56 & 22.20 / 19.65 \\
    \gxpsrnumba & \gxpsr & 21.16  /  17.28 & 21.70  /  18.68 & 21.32  /  19.23 & 21.46  /  19.93 & 19.85 / 18.69 & 22.60 / 20.05 \\
    \onumba & \oname & 21.05  /  17.17 & 21.56  /  18.54 & 21.30  /  19.21 & 21.42  /  19.89 & 19.85 / 18.69 & 22.58 / 20.03 \\
    \pnumba & \pname & 20.80  /  16.92 & 21.74  /  18.72 & 21.38  /  19.29 & 21.46  /  19.93 & 19.75 / 18.59 & 22.74 / 20.19 \\
    \qnumba & \qname & 21.09  /  17.21 & 21.62  /  18.60 & 21.38  /  19.29 & 21.44  /  19.91 & 19.83 / 18.67 & 22.58 / 20.03 \\
    \fbnumba & \fb & 17.29  /  13.41 & 17.26  /  14.24 & 17.13  /  15.04 & 16.43  /  14.90 & 15.53 / 14.37 & 18.33 / 15.78 \\
    \rnumba & \rname & 20.87  /  16.99 & 21.62  /  18.60 & 21.11  /  19.02 & 20.97  /  19.44 & 19.58 / 18.42 & 22.31 / 19.76 \\
    \snumba & \sname & 21.05  /  17.17 & 21.47  /  18.45 & 21.13  /  19.04 & 20.88  /  19.35 & 19.66 / 18.50 & 22.14 / 19.59 \\
    \tnumba & \tname & 21.05  /  17.17 & 21.37  /  18.35 & 21.17  /  19.08 & 21.08  /  19.55 & 19.53 / 18.37 & 22.12 / 19.57 \\
    \unumba & \uname & 18.90  /  15.02 & 19.52  /  16.50 & 19.42  /  17.33 & 19.08  /  17.55 & 18.28 / 17.12 & 20.73 / 18.18 \\
    \hline
    \end{tabular}
    \caption[MeerLICHT magnitude upper limits for the long-term variable sources.]{MeerLICHT upper limits for the long-term variable sources in each band. These are the $5\sigma$ upper limits measured at the position of each source. The first value in each column is the measured upper limit, the second value is the extinction corrected upper limit. {The extinction values were determined using the NASA/IPAC Extragalactic Database Coordinate Transformation and Galactic Extinction Calculator\footnote{\href{https://ned.ipac.caltech.edu/forms/calculator.html}{https://ned.ipac.caltech.edu/forms/calculator.html}}. For the $q$-band extinction we assumed the average extinction of the $g$ and $r$ bands.}}
    \label{tab: VAR supp MLT lims}
\end{table*}



\begin{table*}
    \centering
    \begin{tabular}{llll}
    Catalogue name & Number of sources & Search radius (\asec) & Reference \\
    \hline
    \hline
    GAIA EDR3 & 1721225 & 0.40 & \citet{2020yCat.1350....0G} \\
    GAIA EDR3 x TYCHO & 398 & 0.40 & \citet{2020yCat.1350....0G} \\
    2MASS & 313565 & 0.40 & \citet{2003yCat.2246....0C} \\
    WISE & 99448 & 0.50 & \citet{2012yCat.2311....0C} \\
    AllWISE & 98848 & 0.50 & \citet{2014yCat.2328....0C} \\
    SkyMapper & 269806 & 0.40 & \citet{2018PASA...35...10W} \\
    unWISE & 383183 & 0.40 & \citet{2019ApJS..240...30S} \\
    DENIS & 341161 & 0.50 & \href{http://cds.u-strasbg.fr/denis.html}{http://cds.u-strasbg.fr/denis.html} \\
    2FGL x Radio & 1 & 0.40 & \citet{2015ApJS..217....4S} \\
    YSO candidates & 489 & 0.50 & \citet{2016MNRAS.458.3479M} \\
    Solar-type dwarfs & 3591 & 0.40 & \citet{2016MNRAS.463.4210N} \\
    ASAS-SN variable stars & 295 & 1.00 & \citet{2018MNRAS.477.3145J} \\
    XMM-OM serendipitous 2019 & 46191 & 0.70 & \citet{2012MNRAS.426..903P} \\
    GLADE & 1 & 0.50 & \citet{2018MNRAS.479.2374D} \\
    2MASX ZOA galaxy cat & 3 & 1.50 & \citet{2019MNRAS.482.5167S} \\
    MORX & 143 & 1.00 & \citet{2016PASA...33...52F} \\
    Chandra & 72 & 1.40 & \citet{2010ApJS..189...37E} \\
    Chandra CSC & 114 & 1.40 & \citet{2010ApJS..189...37E} \\
    Swift 2SXPS & 77 & 5.60 & \citet{2020yCat.9058....0E} \\
    ASAS variable stars & 14 & 5.00 & \citet{2002AcA....52..397P} \\
    VISTA & 305142 & 0.40 & \citet{2021yCat.2367....0M} \\
    MGPS-2 & 75 & 2.00 & \citet{2007MNRAS.382..382M} \\
    Tycho & 548 & 0.40 & \citet{tychocat} \\
    Swift UVOT serendipitous & 202777 & 0.50 & \citet{swiftuvotseren} \\
    4XMM serendipitous & 4272 & 1.29 & \citet{xmmseren} \\
    \hline
    \end{tabular}
    \caption[Vizier source catalogues used for source matching.]{Vizier catalogues used for matching sources by position. The number of sources is the total number of sources in the catalogue within a degree of the phase centre of the MeerKAT observations (centred on \gx). The search radius is the maximum matching radius we used for that catalogue, a matching radius of 0\farcs4 indicates that the MeerKAT astrometry is the limiting factor.}
    \label{tab: vizier cats}
\end{table*}


\bsp	
\label{lastpage}
\end{document}